\documentclass[journal,10pt]{IEEEtran}
\usepackage{multirow}
\usepackage{tabu}
\usepackage{makecell}
\usepackage{color}
\usepackage[table]{xcolor}
\usepackage{graphicx}
\usepackage{bm}
\usepackage{balance}
\usepackage{amsmath,amssymb,amsfonts}

\begin{document}

\title{Multi-Perspective Content Delivery Networks Security Framework Using Optimized Unsupervised Anomaly Detection}

\author{\IEEEauthorblockN{Li Yang$^*$, Abdallah Moubayed$^*$,  Abdallah Shami$^*$, Parisa Heidari$^\dagger$, Amine Boukhtouta$^\dagger$, Adel Larabi$^\dagger$, Richard Brunner$^\dagger$, Stere Preda$^\dagger$, and Daniel Migault$^\dagger$}\\
\IEEEauthorblockA{
$^*$Western University, London, Ontario, Canada;
e-mails: \{lyang339, amoubaye, abdallah.shami\}@uwo.ca}
\\
$^\dagger$Ericsson, Montreal, Quebec, Canada;
e-mails: \{parisa.heidari, amine.boukhtouta, adel.larabi, richard.brunner, stere.preda, daniel.migault\}@ericsson.com
}

\markboth{Accepted and to Appear in IEEE Transactions on Network and Service Management}
{}

\maketitle

\begin{abstract}
Content delivery networks (CDNs) provide efficient content distribution over the Internet. CDNs improve the connectivity and efficiency of global communications, but their caching mechanisms may be breached by cyber-attackers. Among the security mechanisms, effective anomaly detection forms an important part of CDN security enhancement. In this work, we propose a multi-perspective unsupervised learning framework for anomaly detection in CDNs. In the proposed framework, a multi-perspective feature engineering approach, an optimized unsupervised anomaly detection model that utilizes an isolation forest and a Gaussian mixture model, and a multi-perspective validation method, are developed to detect abnormal behaviors in CDNs mainly from the client Internet Protocol (IP) and node perspectives, therefore to identify the denial of service (DoS) and cache pollution attack (CPA) patterns. Experimental results are presented based on the analytics of eight days of real-world CDN log data provided by a major CDN operator. Through experiments, the abnormal contents, compromised nodes, malicious IPs, as well as their corresponding attack types, are identified effectively by the proposed framework and validated by multiple cybersecurity experts. This shows the effectiveness of the proposed method when applied to real-world CDN data.
\end{abstract}

\begin{IEEEkeywords}
Cache Pollution Attacks; DoS Attacks; Anomaly Detection; Content Delivery Networks; Gaussian Mixture Model; Bayesian Optimization.
\end{IEEEkeywords}

\IEEEpeerreviewmaketitle

\section{Introduction}
With the increasing popularity of caching techniques in Internet communications, it is estimated that 71\% of Internet traffic will be delivered through content delivery networks (CDNs) by 2021 \cite{caching}. A CDN is a geographically distributed network of servers that work together to provide fast communications of Internet contents, including hypertext markup language (HTML) pages, JavaScript files, images, audios, videos, etc. \cite{cdn2}. CDNs are developed to improve the process of content delivery through caching mechanisms and multiple edge servers \cite{amine}. As a large-scale CDN service provider, Akamai accounts for approximately 20\% of all web traffic \cite{akamai}.
	
CDNs improve the connectivity and efficiency of global communications, but also introduce vulnerabilities to the connected networks. Caching mechanisms become a major target of cyber-attacks since open proxy caches may be exploited by attackers to transmit malicious traffic and perform various harmful activities,  which causes network congestion, unavailability, or other severe consequences \cite{attacksmain}. 

Cache pollution attacks (CPAs) and denial of service (DoS) attacks are the two major types of cyber-attacks launched on CDNs to cause service unavailability or to degrade the caching service by reducing the cache hit rate and increasing latency \cite {cpados_cdn} \cite{dos_cdn}. In high-rate networks, even a moderate degradation of the cache hit rate, or a moderate increase of latency may result in severe network congestion or a massive amount of additional data transmissions \cite{attacksmain}. CPAs are launched by polluting the cache space with a large number of unpopular or illegitimate contents; therefore, legitimate clients will get many cache misses for popular files, making the caching mechanism ineffective \cite{cpados_cdn}. Similarly, DoS attacks are launched by sending a sudden burst of requests to exhaust the network resources of certain targeted nodes. However, DoS attacks are not necessarily launched by sending requests for unpopular files \cite{dos_cdn}.

To protect a CDN against DoS and CPAs, an effective method is to explore and analyze network access logs for the purpose of abnormal behavior analysis and attack pattern detection in CDNs \cite{fe}. Machine learning (ML) algorithms have been widely used in many anomaly detection problems \cite{t1}-\cite{thesis}. In this paper, we focus on anomaly detection in CDNs by analyzing 169 gigabytes (GB) of unlabeled real-world CDN access log data provided by a major CDN operator. This work aims to detect DoS attacks and CPAs based on the behaviors of abnormal network entities, including the malicious Internet Protocol (IP) addresses, abnormal contents, and compromised nodes, through the analysis of access logs. 
The proposed anomaly detection framework consists of a multi-perspective feature engineering, an unsupervised ML model built with optimized isolation forest (iForest) \cite{if} and Gaussian mixture models (GMM) \cite{gmm}, and a multi-perspective result validation method. The proposed work can be considered a labeling technique on unlabeled CDN log data for anomaly detection use cases.

On the other hand, as the data is completely unlabeled, multiple experts from Ericsson Inc. were involved in the learning phase to help construct effective ML models, which is a standard data learning process named human-in-the-loop (HITL). HITL is the process of creating ML models by leveraging the power of both machine and human intelligence \cite{hitl1} \cite{hitl2}. Labeling massive amounts of data usually need HITL to obtain accurate labels for unlabeled data in the unsupervised learning process, as ML models themselves are often unable to determine true labels by themselves \cite{hitl2}.
Therefore, the HITL process is included in the proposed framework to ensure accurate anomaly detection on the unlabeled dataset. HITL in the proposed framework mainly includes the attack pattern \& feature analysis, ML result analysis, and final result validation. 

Detecting potential cyber-attacks and affected abnormal network entities can also trigger other network defense and mitigation mechanisms, such as blacklisting malicious client IPs, isolating compromised nodes, and removing abnormal cached contents out of the cache space \cite{attacksmain}. Thus, CDNs can recover from cyber-attacks or be prevented from potential attacks with the help of effective network anomaly detection techniques.

This paper makes the following contributions:
\begin{enumerate}
\item It summarizes the potential patterns and characteristics of DoS and CPA attacks to assist with anomaly detection in CDNs;
\item It proposes a comprehensive network feature engineering model that generates features from multiple perspectives, including content, client IP, service provider, and account-offering perspectives;
\item It proposes an optimized unsupervised anomaly detection model utilizing iForest, GMM, and Bayesian optimization (BO), to detect cyber-attacks and affected network entities effectively;
\item It proposes a multi-perspective result validation technique that can effectively reduce the false alarm rate and improve the detection rate of unsupervised CDN anomaly detection models. 
\end{enumerate}

This paper is organized as follows: Section II provides an overview of CDNs and potential cyber-attacks. Section III presents the related works regarding network anomaly detection. Section IV discusses the proposed anomaly detection model in detail, including the model framework, feature engineering, utilized algorithms, and validation procedures.  Section V presents and discusses the experimental results. Section VI discusses the open issues and practical usage of the proposed framework. Section VI concludes the paper.

\section{Problem Statement}
\begin{figure*}
     \centering
     \includegraphics[width=14.5cm]{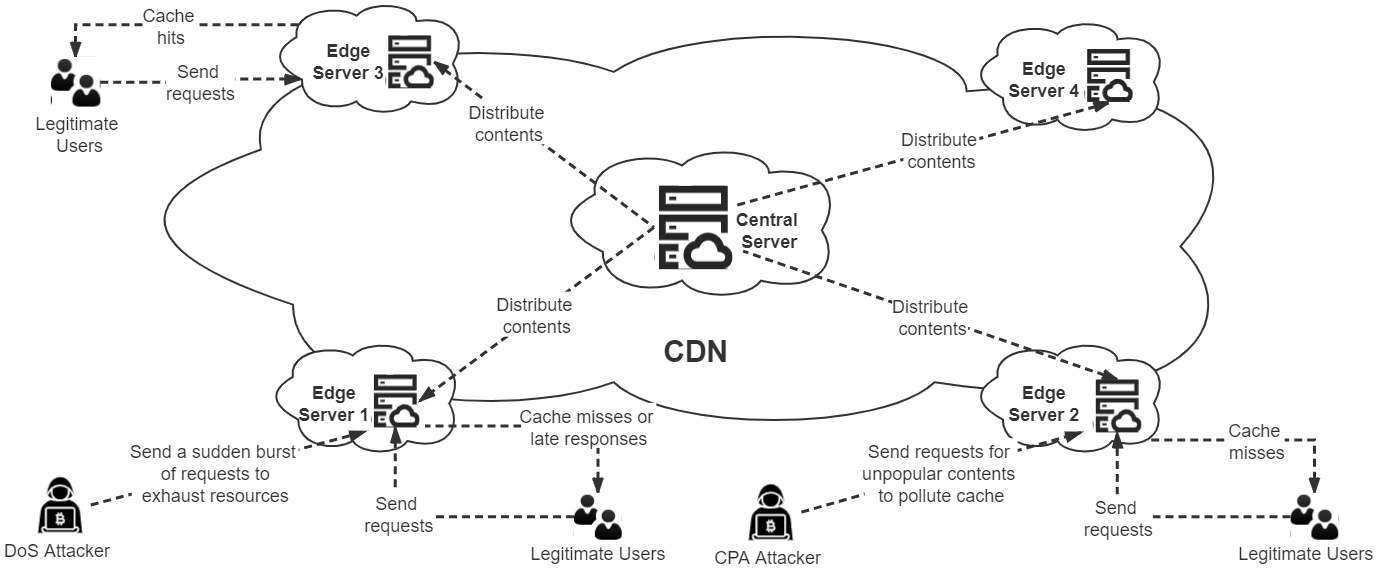}
     \caption{An overview of CDN and cyber-attack scenarios. } \label{CDN}
\end{figure*}

\subsection{CDN Overview}
In traditional Internet, the same contents are required to be transmitted from servers to clients repeatedly. With the rapidly increasing demand for large-scale content distribution, content networks, or caching networks, have been developed to improve content delivery efficiency \cite{cpa_severe} \cite{datacaching}. In content networks, contents can be cached in servers to serve future requests \cite{icnc1}. As a common type of content network and an effective solution for large-scale content delivery, content delivery networks (CDNs) have been widely deployed in modern networks \cite{icnc1}. As the most important strategy of caching networks and CDNs, caching is the process of storing copies of content in temporary storage locations or named caches, which can largely reduce latency, enabling fast access to websites or applications \cite{caching} \cite{datacaching}. Through caching mechanisms, CDNs can cache content in edge servers that are closer to end-users than the central server, making it more efficient to deliver web contents \cite{cdn2}. An overview of CDN is shown in Fig. \ref{CDN}.

CDN works in the following procedures \cite{cdn2}: 
\begin{enumerate}
\item When a CDN receives a request for a content from a client, this request is routed to the closest edge server to the client.
\item The closest edge server will fetch the content from the central server that has this content.
\item The edge server responds to the client with the requested content.
\item A copy of the requested content is stored in the edge server as the caching process for future requests. The cached content will be retained in the cache space if the end-users keep requesting the same content.
\end{enumerate}

If the content requested by an end-user has been saved in the cache space of the edge server, the content will be loaded at a fast speed, so-called a cache hit. In contrast, if the content has not been saved in the cache space, a cache miss will occur, and the edge server will pass the request to the central server to fetch the content and save it in the cache space to reduce the future request processing speed. Thus, CDNs can handle the rapidly growing volumes of network traffic and Internet content with low latency through edge servers and caching mechanisms. Moreover, CDNs can mitigate DDoS attacks because the traffic can be dispersed to multiple edge servers to keep responding to users’ requests, making it difficult for cyber-attackers to paralyze the entire network \cite{protection}.

On the other hand, the research works for CDN anomaly detection are very limited, because CDN operators (e.g., Akamai, Limelight) often keep CDN traffic data private due to liability implications \cite{cdn_no_work}. Most research works of anomaly detection in caching networks are for information-centric networks (ICNs) \cite{attacks2}. 
Information-Centric Networking (ICN), alternatively known as Named Data Networking (NDN) or Content-Centric Networking (CCN), is a future Internet architecture that can be regarded as an improved version of CDNs to provide large-scale content delivery with less resource footprint and system complexity \cite{icnc1} \cite{icnc_improved}. Caching is an essential component in both ICNs and CDNs, since they both aim to provide efficient content delivery through caching mechanisms \cite{caching} \cite{icnc1}. Thus, ICNs and CDNs are both vulnerable to the cyber-attacks that aim to disrupt caching services, like DoS attacks and CPAs. The main difference between ICNs and CDNs is that ICNs assign a unique name for each content as its identifier to replace IP addresses for content delivery, while IP address information is required in CDNs \cite{icnc1} \cite{icnc2}. 

\subsection{Service Targeting Attacks in CDN}
However, the use of caching mechanisms introduces several vulnerabilities to caching networks, especially CDNs \cite{datacaching}. Several service targeting attacks can be carried out on CDNs to disrupt their services by exploiting the caching mechanisms of CDNs. 

Firstly, if a cache misses occurs, even though the content can be an unpopular or illegitimate file, the edge server close to the end-user will fetch this content from the central server and store it in the cache space. However, due to the memory constraints of server machines, a cache can only save a limited number of contents \cite{cpdos}. Based on this caching strategy, cyber-attackers can send a large or moderate number of requests for unpopular contents to occupy the cache space of certain edge servers. This attack is called a cache pollution attack (CPA) that aims to pollute the cache space of certain edge servers \cite{cpa_severe}. CPAs are one of the most severe threats on emerging content networks, including ICNs and CDNs, because the caching mechanisms have made it easier for cyber-attackers to corrupt cache by requesting unpopular or invalid contents \cite {cpados_cdn} \cite{icnc_improved}. The performance and services of caching networks can be significantly degraded by CPAs. After being attacked by CPAs, the cache of compromised nodes is filled with unpopular contents, resulting in a high cache miss rate and latency for popular contents requested by legitimate users \cite{cpa_severe}. Due to the latency issues caused by CPAs, CPAs are especially damaging for the transmission of latency-sensitive contents, like live-streaming contents \cite{cpa_latency}. A CPA scenario for CDNs is shown in Fig. \ref{CDN}.

CPAs can be classified as locality-disruption attacks (LDAs) and false-locality attacks (FLAs) based on their behaviors \cite{attacksmain}. LDAs are launched by sending a moderate number of requests for a large number of low-popularity contents to occupy the cache space that should belong to popular contents, therefore degrading the locality of cache files and cache efficiency. On the other hand, FLAs are launched by repeatedly sending a large number of requests for a few targeted low-popularity contents to constantly refresh these polluted contents and occupy certain areas of the cache space, thereby degrading the cache hit rate of legitimate requests.

Secondly, since CDNs respond to every incoming request with Internet content in chronological order, cyber-attackers can send a sudden burst of requests to certain nodes to overwhelm these edge servers, named denial of service (DoS) attacks \cite{dos_cdn}. DoS attacks are the most common attack that can stop the entire network from functioning shortly or indefinitely to prevent legitimate clients from accessing content \cite{dos_cdn}.

Unlike ICNs that are vulnerable to a special type of DoS attack, named Interest flooding attacks (IFAs), due to its Pending Interest Table (PIT) strategy for content caching, common DoS attacks launched in CDNs are conventional Hypertext Transfer Protocol (HTTP) flooding attacks \cite{dos2}. These DoS attacks intend to exhaust the network and memory resources of CDN nodes, so that the requests from legitimate users may be delayed or their requested contents cannot be cached by these affected nodes, causing cache misses or late responses \cite{protection}. A DoS scenario for CDNs is shown in Fig. \ref{CDN}. 

As a special case of DoS attacks, distributed DoS (DDoS) attacks are carried out by utilizing multiple compromised devices, named bots, to send a sudden burst of requests together to certain nodes, while DoS attacks can be launched by a single machine. It is more difficult to detect DDoS attacks than DoS attacks since they are carried out from multiple locations instead of a single origin \cite{amine} \cite{dos_cdn}.

As CPAs and DoS attacks are two common types of service targeting attacks that pose severe threats to CDNs to make caching mechanisms, the core of CDNs, unavailable to legitimate users, the major purpose of this work is to protect CDNs against these two attacks.

\subsection{Anomaly Detection in CDN}
CDNs themselves can mitigate DoS and DDoS attacks, since they use highly distributed edge servers that can disperse the traffic to avoid the entire system breakdown \cite{protection}. However, latency will still be increased due to the unavailability of edge nodes caused by attacks. Additionally, if DoS attacks cannot be detected and compromised nodes cannot recover, more CDN edge servers and clients will be affected by the attacks, and the entire system will fail eventually. Certain conventional mechanisms, like firewalls and filtering techniques, can mitigate DoS attacks by limiting the amount of traffic entering the CDN. However, DoS attacks and crowd events have similar patterns, making it difficult for conventional mechanisms to filter only the attacks \cite{amine}. Thus, there is a need to develop an anomaly detection system to distinguish them. On the other hand, compared to DoS attacks, CPAs are more flexible, stealthy, and challenging. CPAs can be launched by sending a moderate number of requests for certain unpopular contents to retain them in the cache through the normal content delivery process, making it difficult for conventional mechanisms to defend against CPAs. This emphasizes the importance of anomaly detection system development.

CDNs usually have firewalls and authentication mechanisms as their first layer of defense. Anomaly detection systems can be incorporated into CDNs as the second layer of defense to identify attacks that have breached the first layer of defense. Identifying those attacks is crucial because they may be more malicious than the attacks that are blocked by the first layer of defense \cite{treeme}. Additionally, anomaly detection systems can adapt to the changing patterns of cyber-attacks by analyzing the continuously-generated CDN log data that reflect the current network states. Thus, the work aims to propose an anomaly detection system to detect service targeting attacks (\textit{i.e.}, CPAs and DoS attacks) and abnormal network entities from multiple perspectives to secure CDNs.

\section{Related Work}
\subsection{CPA \& DoS Attack Detection}

Service targeting attacks, including CPAs and DoS attacks, pose a severe threat to caching networks, such as CDNs and ICNs \cite{attacksmain} \cite{datacaching}. As the leading CDN service provider, Akamai Technologies reported in 2019 that more than 800 types of DoS attacks were found in financial services industries \cite{dos_report}. These DoS attacks can exploit vulnerabilities towards the websites and disturb the financial services, causing severe financial losses. Due to the lack of serious scrutiny for cached contents, real-world CDNs are vulnerable to multiple caching-related attacks, like CPAs, which have been recorded in the “Vulnerability Notes Database” \cite{cpa_cdn}. Moreover, Nguyen \textit{et al.} \cite{cpdos} proposed a new caching attack, Cache-Poisoned Denial-of-Service (CPDoS), that targets CDNs and other vulnerable caching systems. CPDoS attacks combine the ideas of CPA and DoS attacks by sending a large number of requests with malicious headers to pollute cache space and paralyze victim websites.

Several research works have focused on CPA detection. Conti \textit{et al.} \cite{attacks2} conducted experiments to prove that CPAs are a realistic threat to caching networks and proposed a lightweight detection technique to detect CPAs accurately. However, the experiments were conducted on a simulated network instead of a real-world network. Xie \textit{et al.} \cite{cpa_p1} proposed a novel method named CacheShield to avoid storing the low-popularity contents (less than the popularity threshold) in servers’ caches. However, this strategy can also reject certain legitimate contents, and is inefficient for gradually enhanced attacks. Karami \textit{et al.} \cite{cpa_p2} proposed an Adaptive Neuro-Fuzzy Inference System (ANFIS) to mitigate CPAs in ICNs. In ANFIS, every content will be given a grade to measure its goodness, and the system will replace low-grade contents in caches with high-grade contents based on a goodness threshold to mitigate cache pollution. However, it has high overhead costs. On the other hand, the above CPA detection methods are all based on threshold mechanisms, but they lack adaptability to complex and changeable network environments.

Many recent research works have considered DoS attack prevention and detection. Rahman \textit{et al.} \cite{dos_cdn} proposed a distributed virtual honeypot method to mitigate DoS attacks in CDNs. This method can maintain smooth content delivery in CDN edge servers, but cannot protect the main server. Moubayed \textit{et al.}  \cite{abdns1} \cite{abdns2} proposed an ensemble learning classifier to effectively detect several types of domain name system (DNS) attacks, including cache poisoning and DoS attacks. The proposed method achieves high accuracy but does not analyze the detection results with related features to summarize attack patterns for future anomaly detection. Kumar \textit{et al.}  \cite{absdp1} \cite{absdp2} proposed a security framework based on the software-defined perimeter (SDP) to protect modern networks from being breached by DoS attacks. Certain DoS attacks can be effectively defended or prevented by the proposed method, but it lacks the capacity to detect the malicious attacks that have already breached the networks.

\subsection{Abnormal IP \& Node Detection}
Detecting abnormal network entities, including abnormal client IP addresses and nodes, is important for CDN protection. Several recent works have extracted network features from the client IP point of view for anomaly detection. Lee \textit{et al.} \cite{dos2} proposed CDN request routing and site allocation algorithms to distinguish between the requests from DoS attackers and legitimate users in CDNs, but they did not consider other attack types. Chiba \textit{et al.} \cite{ipdetect1} proposed a novel method that can effectively extract features from the structures of IP addresses and applied the support vector machine (SVM) model to detect malicious websites. Pinto \textit{et al.} \cite{ipdetect2} presented and utilized SVM on the network traffic data to identify malicious IP addresses and achieved the cross-validation accuracy of 83\% to 95\%. Fiadino \textit{et al.} \cite{ipdetect3} used the HTTP flow data collected from a primary European Internet service provider to detect traffic anomalies in CDNs. However, only detecting numerically abnormal traffic is insufficient to identify cyber-attacks accurately. 

Detecting compromised nodes is also a critical process to ensure a quick recovery and reliable functioning of networks. La \textit{et al.} \cite{nodedetect1} proposed a misbehavior node detection algorithm using a weighted-link method to secure a hierarchical sensor network. Berjab \textit{et al.} \cite{nodedetect2} proposed a novel framework based on observed spatiotemporal (ST) and multivariate-attribute (MVA) sensor correlations to detect abnormal nodes in wireless sensor networks (WSNs). Pandey \textit{et al.} \cite{nodedetect3} proposed an innovative method that uses intrusion detection system (IDS) agents to identify compromised nodes in WSNs according to their behavior, and the proposed method shows efficiency in small networks. However, the above methods lack a root cause analysis to find what caused the abnormal nodes for the purpose of future intrusion prevention.

\subsection{Research Contributions}
Many of the research works presented in this section are promising and have achieved good outcomes. However, most of them use software like the network simulator version 3 (NS-3) to build a simulated network for data collection, which may be biased or noisy. In our work, a recent real-world CDN access log dataset collected from a major operator is used to show the effectiveness of applying our proposed anomaly detection model to real-world networks. 

On the other hand, most existing research works only detect anomalies from a single perspective, like the client IP or the service provider perspective. This is often insufficient to validate whether the detected abnormal network behavior is due to malicious cyber-attacks or legitimate network events. Thus, many false alarms may be returned. Caltagirone \textit{et al.} \cite{diamond} proposed a pivoting intrusion analysis model named “diamond” that uses the communication information between compromised nodes and malicious IP addresses to reveal the detail of attackers, but it is only a theoretical model. On the other hand, our proposed method extracts more CDN attributes/features and conducts anomaly detection from several different perspectives, and then performs a multi-perspective analysis to validate anomaly detection results for the purpose of false alarm reduction. 

Moreover, many recent research works treat ML models as black-box methods and do not analyze how the detected anomalies and corresponding features can reflect a specific type of cyber-attack. In our proposed method, the behaviors and characteristics of CPA and DoS attacks from multiple perspectives are summarized and analyzed together with the ML-based anomaly detection results to perform a root cause analysis and find which type of cyber-attack or event causes the anomalies. Ultimately, both the abnormal network entities and their corresponding cyber-attack types will be detected effectively.

\section{Proposed Anomaly Detection Framework}
\subsection{System Overview and Deployment}
\begin{figure*}
     \centering
     \includegraphics[width=13.5cm]{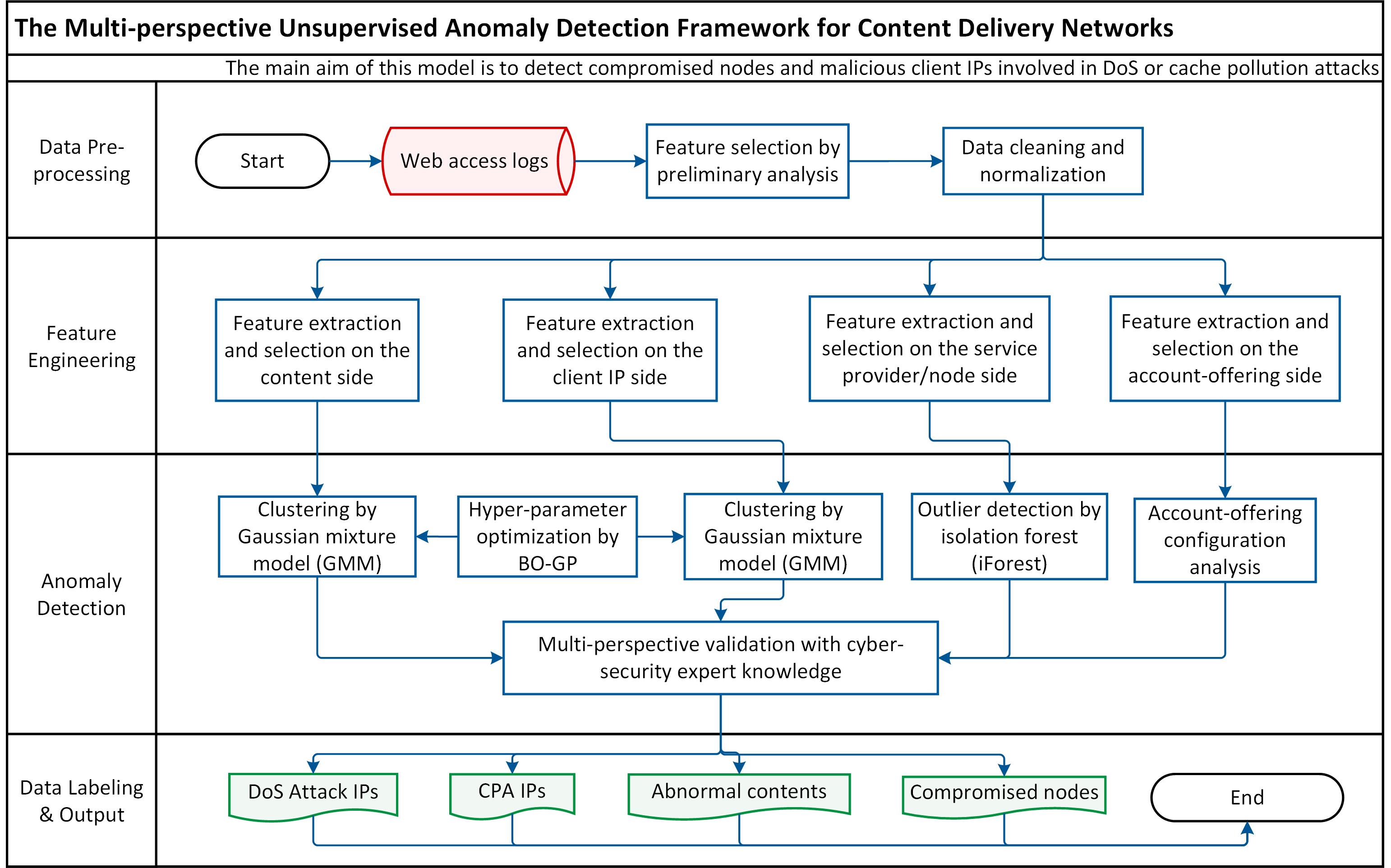}
     \caption{The framework of the proposed anomaly detection approach. } \label{flow}
\end{figure*}

The proposed method aims to characterize abnormal events and separate them from normal network events based on the analytics of CDN access log data from different perspectives, including content, client IP, account-offering, and node perspectives. Fig. \ref{flow} depicts the framework of the proposed anomaly detection model. The overall architecture of the proposed system is divided into four parts: data pre-processing, feature engineering, anomaly detection, and data labeling. At the first stage, the raw access log data is pre-processed and cleaned to generate a sanitized dataset. A comprehensive feature engineering method is then implemented to extract the datasets that can effectively reflect the behaviors of network attacks from different perspectives. Next, the extracted datasets are trained by an optimized unsupervised anomaly detection model based on GMM, iForest, and BO to discern between abnormal and normal data patterns. At the last stage, a multi-perspective validation analysis is conducted to reduce the errors in the anomaly detection results. Ultimately, abnormal network entities, including malicious IPs, abnormal contents, and compromised nodes, as well as their corresponding cyber-attack types, can be effectively detected to secure CDNs.

\begin{figure}
     \centering
     \includegraphics[width=8.8cm]{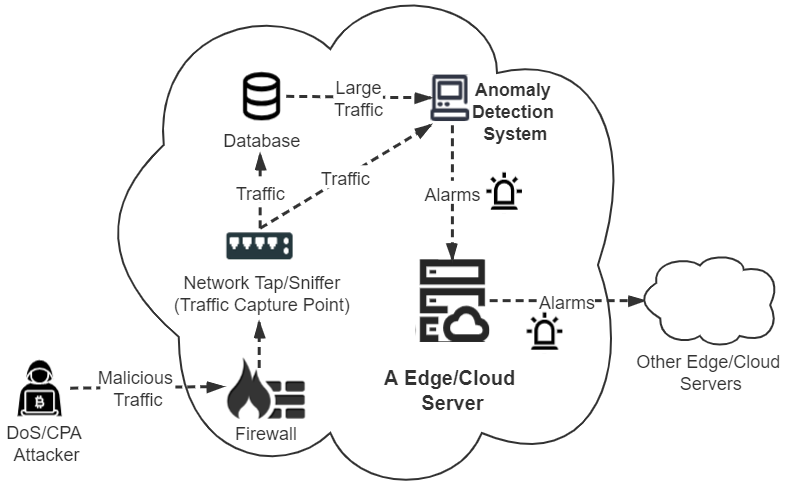}
     \caption{The deployment of the proposed anomaly detection system. } \label{deploy}
\end{figure}

For the deployment in CDNs, the proposed anomaly detection system can be placed in both the central server and edge servers, as shown in Fig. \ref{deploy}. In edge servers, the proposed system can keep monitoring the network traffic to detect abnormal network entities and send alarms to the central server as soon as an attack occurs; hence, the central server can notice other edge servers and make corresponding countermeasures. When placed in the central server, the proposed system can have a comprehensive view of the operation of the entire network, and can protect the central server when certain edge servers have been exploited by attackers to breach the central server. Specifically, in each edge or cloud server, all the network traffic that is not stopped by the first layer of defense (\textit{e.g.}, firewalls) will be captured by network taps or sniffers, and then analyzed by the proposed anomaly detection system \cite{mth}. The large traffic can also be stored in a database for comprehensive analysis by the proposed anomaly detection system. If an attack is detected in an edge or cloud server, all the edge and cloud servers will receive an alarm. As such, the network administrators in the central server and edge servers can make corresponding countermeasures to stop current attacks and prevent future attacks.

\section{Problem Statement}

\subsection{Data Acquisition and Preprocessing}
For the purpose of network anomaly detection, data acquisition is the first phase of any ML-based model framework. Due to the complexity of modern network configurations, numerous network fields are often recorded in network log data to reflect network states and characteristics. For instance, the authors in \cite{248features} provided 248 unique network features that can be collected from the network packets transmitted between clients and servers. However, considering hundreds of network features for anomaly detection is often unrealistic due to limited budgets and resources. Additionally, a large number of features that include many irrelevant features may introduce noises, which have an adverse impact on the performance of ML models. 

In this paper, we focus on anomaly detection in CDN access logs. In network communications, access logs record the information of all requests for the contents that users have requested from web servers. The information includes client IP addresses, timestamps, protocol, user-agent information, uniform resource locator (URL) of content, etc. \cite{accesslog}. Analyzing access logs can help us recognize the states and behaviors of a network in various time periods. Access logs are often recorded in a combined log format that contains multiple network fields for each request, as described in \cite{accesslog}. The following is an example of a record in general access logs: 

\textit{127.0.0.1 - - [12/Dec/2016:04:54:20 -4000] “GET /support.html HTTP/1.1” 200 11179 - “Mozilla/5.0(compatible;Googlebot/2.1;+http://www.g oogle.com/bot.html)” …}

The access log data used in our paper is a general network access log acquired from real-world web servers. In this 169 GB access log data provided by the CDN operator, there are more than 0.4 billion requests/samples and 30 fields that characterize the behavior of each request. This CDN log data contains data samples with various characteristics and probability distributions, enabling the detection of different types of cyber-attacks. 

Although 30 network fields are available in the CDN access logs, only part of them are useful for anomaly detection, and other fields are irrelevant or for other uses.
After preliminary analysis on the CDN dataset and potential abnormal behaviors, 12 fields that might be helpful for anomaly detection are selected, including IP address, timestamp, HTTP method, status code, bytes returned, request delivery time, service type, cache hit indicator, node name, account-offering, content-URL, and content type. These features are also the standard access log features described in \cite{accesslog}. The description of the preliminarily selected network fields is shown in Table \ref{raw}.
Other fields that do not have a direct impact on cyber-attack detection, such as protocol, network name, referrer string, are removed from the feature set, because they have either the same values or empty values for almost all the requests. To ensure accurate feature selection, HITL is used for feature validation. Thus, this preliminary feature selection process is validated by multiple cybersecurity experts and industrial partner security network engineers. The experts are from multiple organizations, including third and disinterested parties. All the experts involved in this work can guarantee their professionalism, objectivity, and impartiality in the analysis procedures.

\begin{table}[]
\caption{Description of Selected Network Log Attributes}
\centering
\setlength\extrarowheight{1pt}

\scalebox{0.85}{
\begin{tabular}{|>{\centering\arraybackslash}p{5.5em}|p{27em}|}

\hline
\textbf{Feature}               & \textbf{Description}                                                                                                                                        \\ \hline
IP                    & Client IP address                                                                                                                                  \\ \hline
Timestamp             & \makecell[l]{Request start time in format:  “{[}dd/mmm/yyyy:hh:mm:ss -zzzz{]}”}                                                                                  \\ \hline
HTTP   method         & HTTP request method, \textit{e.g.},   GET, POST, etc.                                                                                                       \\ \hline
Status   code         & HTTP status code: 2xx indicates a successful   response; 3xx indicates a redirection; 4xx indicates a client error; 5xx indicates   a server error \\ \hline
Bytes                 & Bytes returned over the network without headers after   a request                                                                                  \\ \hline
Delivery   time       & Duration from the beginning until the end of a request   and all bytes are delivered, in milliseconds                                              \\ \hline
Service   type        & The type of service, \textit{e.g.}, static, live   streaming, progressive download, etc.                                                                    \\ \hline
Cache   hit indicator & Service cache hit/miss indicator, hit or miss                                                                                                      \\ \hline
Node                  & Node name, representing a service provider                                                                                                         \\ \hline
Account   offering    & The network account offering name (\textit{e.g.},   streaming-1, static-2, etc.), indicating a service accessed from the IP   space                 \\ \hline
Content-URL           & The content part of a URL, indicating unique content                                                                                              \\ \hline
Content   type        & The type of the content in each request, \textit{e.g.},   image, video, audio, text, etc.                                                                   \\ \hline
\end{tabular}
}
\label{raw}%
\end{table}

After obtaining the reduced raw data, several steps are completed to clean and pre-process the data. 

In the first step, apparent noise or error data, mainly the request samples with many empty fields, are discarded. Then, certain features, like the timestamp, are formatted for easier comparison or calculation. Moreover, string features, like the service category (dynamic or static) and cache hit indicator (hit or miss), are converted to binary or numerical features for easier calculation. 

Data cleaning is then followed by data normalization. Data normalization is beneficial when features have different ranges since features with larger ranges are often considered more important than smaller range features in ML model training, which can cause misleading results. Therefore, the datasets are normalized by the min-max normalization method to be in the range of 0 to 1 \cite{treeme}. The normalized value of each feature value, $x_n$, is denoted by,
\begin{equation}
x_{n}=\frac{x-min}{max-min},
\end{equation}%
where $x$ is the original feature value, $min$ and $max$ are the minimum and maximum values of each original feature.

Min-max normalization is chosen in the proposed framework due to two main reasons \cite{scaler}:
\begin{enumerate}
\item Min-max scaling can transform all the features to have the same range of [0, 1], making it easier for ML models to process the dataset and for people to compare the results; while many other normalization methods cannot obtain the exact same range of features;
\item Compared to other methods, like standard scaler and power transformer, the min-max scaler is more sensitive to outliers, enabling the ML algorithms to detect anomalies more accurately, as the purpose of this work is to identify anomalies and attacks.
\end{enumerate}

The use of min-max normalization can improve the accuracy of ML models and reduce the difficulty of result analysis. Nevertheless, the performance of using different normalization methods is often very similar \cite{scaler}. Additionally, as normalization is a small part of ML pipelines, it does not have a significant impact on the final anomaly detection results. Other procedures, including feature engineering, unsupervised learning model development, and multi-perspective validation, are more important.

\subsection{Feature Engineering}
Although there are 12 features for each request in the initial dataset, it is difficult to identify abnormal network entities and cyber-attacks using only the original request dataset since a single request is often insufficient to reflect network anomalies or attacks. Thus, other features that can effectively reflect abnormal network behaviors should be extracted or generated for the purpose of anomaly detection.

Feature engineering aims to obtain useful features from the log traces collected from CDN deployment based on domain knowledge \cite{amine}. In this paper, a multi-perspective feature engineering method that extracts features from four main perspectives: content, client IP, service provider (node), and account-offering, is proposed to obtain the dedicated datasets for different purposes. The generated datasets extracted from different perspectives can be processed separately and then analyzed together to detect abnormal network events more accurately than any single perspective. In the proposed framework, detecting malicious client IPs and compromised nodes is the main objective; the information collected from the content and account-offering perspectives is the supporting information for more accurate abnormal IP and node detection. To detect CPAs and DoS attacks, their main characteristics reflected from the four considered perspectives are summarized below.  

The content perspective is to monitor the properties of the requested and cached contents. From the content perspective, it is beneficial to identify the abnormal contents that might be used by attackers to launch cyber-attacks. The abnormal traffic that characterizes a CPA is a sudden burst or a periodic sending of requests for low popularity contents. Additionally, most CPAs are used to breach a few targeted nodes, so for each abnormal content, the number of requests sent to each targeted node should also be large. On the other hand, if a content gets a large number of requests from many different IPs, there might be a crowd event or a distributed denial of service (DDoS) attack. 

The service provider perspective is meant to monitor the edge servers or nodes that receive requests and transmit contents through CDNs. The cache hit rates and data transfer rates of compromised nodes are often degraded due to uncached legitimate contents and network unavailability/congestion caused by CPAs or DoS attacks. The average content popularity of each compromised node is also reduced by CPAs since the node’s cache space will be occupied by unpopular files. The affected nodes should be identified and isolated as soon as an intrusion occurs so that other legitimate nodes can avoid communicating with the compromised nodes until recovery.

The client IP perspective is meant to monitor clients that send requests for the contents provided by servers. Considering the client IP perspective enables us to detect potential malicious clients, allowing us to stop or prevent cyber-attacks by blocking the requests from these malicious IP addresses. Detecting abnormal client IPs is crucial because these IP addresses represent the origins of cyber-attackers. Attackers that are launching different types of attacks exhibit different behaviors. For CPA attackers, they send high-frequency requests for low popularity contents, while DoS attackers may send a sudden burst of requests for files with any popularity. Additionally, for the two types of CPA attackers, LDA attackers send requests for a large number of unpopular contents, while FLA attackers only send requests for a few targeted contents, but the number of requests sent for each content (request per content ratio) is often large. 

The account-offering (AO) perspective indicates the streaming source that provides a specific type of service, such as static, live streaming, and progressive download content distribution. An AO’s configuration also determines the behavior of its serviced client IPs. If certain IPs behave abnormally and do not match the AO configuration patterns, they have a high probability of being anomalies. Thus, the analysis from the AO perspective can help us validate whether the anomaly detection results are real attacks or false alarms. 

The extracted features from four perspectives are summarized in Tables \ref{contentFE} - \ref{aoFE}. After attack pattern analysis in Section II-B, we found that only some of the features have a direct impact on the detection of CPAs and DoS attacks. Considering more features may cause misleading results or additional computational time. Thus, the most important features that can directly reflect the attack patterns are selected in the proposed method, based on the potential feature patterns/characteristics of CPAs and DoS attacks summarized in Tables \ref{cpapattern} and \ref{dospattern}. The selected features can reflect most scenarios of a CDN that is under CPA or DoS attacks. 

Considering the original features in Table \ref{raw}, except for the HTTP method, all other raw features in Table \ref{raw} have been used to create the final features in Tables \ref{aoFE} - \ref{dospattern}. Specifically, “IP”, “Node”, and “Content-URL” are used to extract datasets from the IP, node, and content perspectives, “Timestamp” is used to calculate the average request interval of client IPs, “Status code” is used to calculate the request error rate of nodes, “Bytes” and “Delivery time” are used to calculate the data transfer rate, “Cache hit indicator” is used to calculate the cache hit rate of IPs and nodes, “Account-offering”, “Service type”, and “Content type” are used to determine the behaviors of account offerings for result validation purposes. Therefore, the fields selected by the cybersecurity experts do not include any noise.  “HTTP method” is removed because DoS and CPAs can be launched using any HTTP method, whether they use GET, POST, or any other HTTP methods; hence, it is irrelevant for DoS \& CPA detection. Although the “HTTP method” is irrelevant for CPA \& DoS attack detection, it can still be used for other tasks, like the detection of other types of attacks. 

This feature extraction and selection process can be automated as a general feature engineering method by summarizing the potential patterns of certain types of cyber-attacks and selecting the features that can reflect the attack patterns, as discussed in this subsection. Through this process, the proposed multi-perspective feature engineering method can extract and select the core features to reflect CPA and DoS attacks for the anomaly detection model development presented in the next subsection.

\begin{table}[]
\caption{Description of Extracted Features from the Content Perspective}
\centering
\setlength\extrarowheight{1pt}

\scalebox{0.85}{
\begin{tabular}{|>{\centering\arraybackslash}p{4.5em}|>{\centering\arraybackslash}p{5em}|p{18em}|}

\hline
\textbf{Perspective}              & \textbf{Feature }                 & \textbf{Description}                                                                                                                         \\ \hline
\multirow{10}{*}{Content} & Number of requests     & The total number of requests for per content                                                                                        \\ \cline{2-3} 
                         & Popularity               & The popularity of per content, represented by the normalized number of IPs that sent requests to per content                        \\ \cline{2-3} 
                         & Cache hit rate         & The number of cache hits divided by the total number of requests for per content                                                    \\ \cline{2-3} 
                         & Request per IP ratio   & The ratio of the total number of requests sent for per content to the total number of IPs which sent requests for per content       \\ \cline{2-3} 
                         & Request per node ratio & The ratio of the total number of requests sent for per content to the total number of nodes which received requests for per content \\ \hline
\end{tabular}
}
\label{contentFE}%
\end{table}

\begin{table}[]
\caption{Description of Extracted Features from the Node Perspective}
\centering
\setlength\extrarowheight{1pt}

\scalebox{0.85}{
\begin{tabular}{|>{\centering\arraybackslash}p{4.5em}|>{\centering\arraybackslash}p{5em}|p{18em}|}

\hline
\textbf{Perspective}              & \textbf{Feature }                 & \textbf{Description}                                                                                                                         \\ \hline
\multirow{16}{*}{Node} & Cache   hit rate                   & The number of cache hits divided by the total number   of requests received by per node                                 \\ \cline{2-3} 
                      & Cache   hit rate of legitimate IPs & Average cache hit rate of IPs which only requested   for popular contents on per node                                   \\ \cline{2-3} 
                      & Data   transfer rate (MB/s)        & The total bytes returned divided by the total   delivery time for per node                                              \\ \cline{2-3} 
                      & Request   error rate               & The percentage of requests with errors (4xx/5xx status code)   received by per node                                              \\ \cline{2-3} 
                      & Average   request popularity       & Average content popularity of requests received by   per node                                                           \\ \cline{2-3} 
                      & Account-offering   request rate    & The percentage of requests sent through per   account-offering for per node, \textit{e.g.}, “account1: 80\%, account2: 20\%” \\ \hline
\end{tabular}
}
\label{nodeFE}%
\end{table}

\begin{table}[]
\caption{Description of Extracted Features from the Client IP Perspective}
\centering
\setlength\extrarowheight{1pt}

\scalebox{0.85}{
\begin{tabular}{|>{\centering\arraybackslash}p{4.5em}|>{\centering\arraybackslash}p{5em}|p{18em}|}

\hline
\textbf{Perspective}              & \textbf{Feature }                 & \textbf{Description}                                                                                                                         \\ \hline
\multirow{22}{*}{Client   IP} & Number   of requests            & The total number of requests sent by per IP                                                                                 \\ \cline{2-3} 
                              & Average   request interval      & The average time interval between consecutive   requests sent by per IP                                                     \\ \cline{2-3} 
                              & Number   of nodes               & The total number of unique nodes that received   requests from per IP                                                       \\ \cline{2-3} 
                              & Number   of contents            & The total number of unique contents requested by per IP                                                                 \\ \cline{2-3} 
                              & Request   per content ratio     & The ratio of the total number of requests sent by per   IP to the total number of contents requested by per IP              \\ \cline{2-3} 
                              & Request   per node ratio        & The ratio of the total number of requests sent by per   IP to the total number of nodes which received requests from per IP \\ \cline{2-3} 
                              & Average   request popularity    & Average content popularity of requests sent by per IP                                                                       \\ \cline{2-3} 
                              & Cache   hit rate                & The number of cache hits divided by the total number   of requests sent by per IP                                           \\ \cline{2-3} 
                              & Request   error rate            & The percentage of requests with errors (4xx/5xx status code) sent by per IP                                                        \\ \cline{2-3} 
                              & Account-offering   request rate & The percentage of requests are sent through per   account-offering for per IP, \textit{e.g.}, “account1: 80\%, account2: 20\%”       \\ \hline
\end{tabular}
}
\label{ipFE}%
\end{table}

\begin{table}[]
\caption{Description of Extracted Features from the Account-Offering Perspective}
\centering
\setlength\extrarowheight{1pt}

\scalebox{0.85}{
\begin{tabular}{|>{\centering\arraybackslash}p{4.5em}|>{\centering\arraybackslash}p{5em}|p{18em}|}

\hline
\textbf{Perspective}              & \textbf{Feature }                 & \textbf{Description}                                                                                                                         \\ \hline
 & Number   of requests & The total number of requests through per AO                                                        \\ \cline{2-3} 
                                         & Number   of nodes    & The total number of unique nodes that received   requests sent through per AO                           \\ \cline{2-3} 
Account-offering   (AO)                                                             & Service   type       & The type of service provided by per AO, \textit{e.g.},   static, live streaming, progressive download, etc. \\ \cline{2-3}                   & Content   type       & The type of content provided by per AO, \textit{e.g.},   image, video, audio, text, etc.  \\ \cline{2-3}          
& Cache   hit rate     & The number of cache hits divided by the total number of requests sent through per AO \\ \cline{2-3} 

                                         & Request   popularity & Average content popularity of requests sent through   per AO                                       \\ \hline
\end{tabular}
}
\label{aoFE}%
\end{table}

\begin{table}[]
\caption{Potential Patterns of CPAs}
\centering
\setlength\extrarowheight{1pt}

\scalebox{0.85}{
\begin{tabular}{|>{\centering\arraybackslash}p{3em}|>{\centering\arraybackslash}p{4.5em}|>{\centering\arraybackslash}p{13em}|>{\centering\arraybackslash}p{5em}|}

\hline
\textbf{Attack Type}           & \textbf{Perspective}                  & \textbf{Feature}                            & \textbf{Abnormal Patterns }                                                      \\ \hline
\multirow{13}{*}{CPA} & \multirow{4}{*}{Node}        & Cache   hit rate                   & Low                                                                     \\ \cline{3-4} 
                      &                              & Cache   hit rate of legitimate IPs & Low                                                                     \\ \cline{3-4} 
                      &                              & Data   transfer rate (MB/s)        & Low                                                                     \\ \cline{3-4} 
                      &                              & Average   request popularity       & Low                                                                     \\ \cline{2-4} 
                      & \multirow{6}{*}{Client   IP} & Number   of requests               & Large                                                                   \\ \cline{3-4} 
                      &                              & Average   request interval         & Short                                                                   \\ \cline{3-4} 
                      &                              & Number   of nodes                  & Small                                                                   \\ \cline{3-4} 
                      &                              & Request   per content ratio        & LDA: Low   FLA:   High \\ \cline{3-4} 
                      &                              & Average   request popularity       & Low                                                                     \\ \cline{2-4} 
                      & \multirow{4}{*}{Content}     & Popularity                         & Low                                                                     \\ \cline{3-4} 
                      &                              & Request   per IP ratio             & FLA: High LDA: Low      \\ \cline{3-4} 
                      &                              & Request   per node ratio           & High                                                                    \\ \cline{2-4} 
                      & AO             & Request   popularity               & Low                                                                     \\ \hline
\end{tabular}
}
\label{cpapattern}%
\end{table}

\begin{table}[]
\caption{Potential Patterns of DoS Attacks}
\centering
\setlength\extrarowheight{1pt}

\scalebox{0.85}{
\begin{tabular}{|>{\centering\arraybackslash}p{3em}|>{\centering\arraybackslash}p{4.5em}|>{\centering\arraybackslash}p{13em}|>{\centering\arraybackslash}p{5em}|}

\hline
\textbf{Attack Type}           & \textbf{Perspective}                  & \textbf{Feature}                            & \textbf{Abnormal Patterns }                                                      \\ \hline
\multirow{10}{*}{DoS} & \multirow{4}{*}{Node}        & Cache   hit rate                   & Low               \\ \cline{3-4} 
                      &                              & Cache   hit rate of legitimate IPs & Low               \\ \cline{3-4} 
                      &                              & Data   transfer rate (MB/s)        & Low               \\ \cline{3-4} 
                      &                              & Request   error rate               & High              \\ \cline{2-4} 
                      & \multirow{5}{*}{Client   IP} & Number   of requests               & Large             \\ \cline{3-4} 
                      &                              & Average   request interval         & Short             \\ \cline{3-4} 
                      &                              & Number   of nodes                  & Small             \\ \cline{3-4} 
                      &                              & Cache   hit rate                   & Low               \\ \cline{3-4} 
                      &                              & Request   error rate               & High              \\ \cline{2-4} 
                      & AO             & Cache   hit rate                   & Low               \\ \hline
\end{tabular}
}
\label{dospattern}%
\end{table}

\subsection{Unsupervised Anomaly Detection}
\subsubsection{Compromised Node Detection}
The extracted node-based dataset contains the information about 50 different nodes. To detect abnormal or compromised nodes, isolation forest (iForest) \cite{if}, an unsupervised outlier detection algorithm that aims to separate isolated data samples (anomalies) from normal samples, is utilized in the proposed framework. 
The proposed abnormal node detection method has two main steps: 
\begin{enumerate}
\item Use iForest, an outlier detection algorithm, to separate numerically abnormal samples from normal samples. 
\item Analyze the behaviors of each numerically abnormal node, and label the nodes that match the summarized patterns/characteristics of different types of attacks.
\end{enumerate}

IForest is an ensemble learning algorithm constructed with multiple binary search trees, named isolation trees (iTrees) \cite{if2}. Each iTree is constructed by splitting the samples based on feature values. The number of splittings required to isolate a sample indicates its path length (\textit{i.e.}, the number of edges from an iTree's root node to its leaf node). The path length of anomalies is often shorter than normal samples. This is because normal samples are often the majority and in dense areas, making it unlikely for an iTree to isolate them from each other, while anomalies are the opposite. A score is given to each sample based on the path length, so that the outliers that are sparsely distributed and distant from the dense normal samples can be detected \cite{if2}. 

IForest is chosen for abnormal node detection due to the following reasons \cite{if} \cite{if2}:
\begin{enumerate}
\item Unlike many other ML algorithms, iForest performs well on small-scale data to which the node-based dataset belong. This is because iForest uses the short path length of data samples to indicate the anomalies, which can be obtained regardless of data size. 
\item Unlike clustering algorithms, iForest does not require anomalies to have similar characteristics to detect them since it discerns outliers from normal samples based on data density.
\item IForest is computationally efficient because it has a linear time complexity of $O(N)$, a low memory requirement, and parallel execution support.
\item IForest has good interpretability since it uses a tree-structure to make decisions and split data samples.  
\end{enumerate}

To develop an effective ML model for a specific task, hyper-parameter tuning should be implemented to detect the hyper-parameter configuration that can return the optimal architecture of the ML model \cite{hpm1}-\cite{driftmag}. As an important hyper-parameter of iForest, the contamination level determines the proportion of data samples that will be detected as outliers. To build an optimized iForest model, the contamination level is tuned using Bayesian optimization (BO).

BO algorithms are a set of efficient hyper-parameter optimization (HPO) methods that detect the optimal hyper-parameter based on the currently-evaluated results \cite{hpm3}. In BO, a surrogate model is used to fit all the currently-tested samples into the objective function; an acquisition function is then used to locate the next points by considering both the unexplored regions and currently-promising regions in the search space \cite{hpm4}. 

Gaussian process (GP) is a common surrogate model for BO. In GP surrogate models, any finite combination of the random variables follows a Gaussian distribution \cite{hpme}: 
\begin{equation}
p(y | x, D)=N\left(y | \hat{\mu}, \hat{\sigma}^{2}\right),
\end{equation}	
where $D$ is the hyper-parameter configuration space, $y=f(x)$ is the objective function value for each hyper-parameter configuration with its mean as $\hat{\mu}$ and covariance as $\hat{\sigma}^{2}$. 

The BO method using GP (BO-GP) has a time complexity of $O(N^3)$ and a space complexity of $O(N^2)$ \cite{hpme}. BO-GP is inefficient for a large hyper-parameter search space but exhibits great performance on optimizing a small number of continuous or discrete hyper-parameters. Thus, BO-GP is used to optimize the contamination level of the iForest model. The silhouette coefficient \cite{sklearn} \cite{silam}, a distance-based metric that can measure the similarity of normal samples and the difference between numerically normal and anomalous samples, is chosen to be the metric of the iForest model and used as the objective function to be optimized by BO-GP. 

After using the optimized iForest model to detect abnormal nodes, many false positives will be returned since many of the detected compromised nodes do not match the cyber-attack patterns. Mainly, the cache hit rate of the compromised nodes and their serviced legitimate client IPs should be reduced due to attacks. The data transfer rate and average popularity of the abnormal nodes should also be low due to network congestions and the filled cache space occupied by unpopular files, respectively. On the other hand, although certain other isolated data points, like the nodes that received a very small number of requests or returned a very high data transfer rate, are numerically different from most normal nodes, they are unlikely to be under CPA or DoS attacks. Therefore, the patterns of the detected anomalous nodes are compared with the behaviors of potentially compromised nodes summarized in Tables \ref{cpapattern} \& \ref{dospattern}. Ultimately, the affected nodes that match the abnormal patterns are preliminarily labeled abnormal. 

\subsubsection{Abnormal Content \& Client IP Detection}
Unlike the node-based dataset that only contains 50 unique nodes, both the content-based datasets and IP-based datasets have more than one million unique contents or unique IPs. For large-scale data, only using a binary outlier detection method (\textit{e.g.}, iForest)  is insufficient to identify attack samples, because using only two categories cannot describe various normal and abnormal patterns, and numerical anomalies may not be the true attack samples since cyber-attacks have their own specific patterns or characteristics, as described in Section IV-C. 

For large-sized data, clustering algorithms would be more effective in identifying abnormal contents and IP addresses since they can obtain multiple numerically abnormal clusters, enabling us to determine the true anomalies by comparing the characteristics of the abnormal clusters with the specific attack patterns. Clustering algorithms are a set of unsupervised learning models that aim to group data points into different clusters. Data samples in the same cluster should have similar patterns or properties, while those in different clusters should have different patterns \cite{clusteram}. 

The proposed abnormal IP and content detection method has two main steps: 
\begin{enumerate}
\item Use a clustering algorithm to group the client IPs or contents into a sufficient number of clusters. 
\item Analyze the characteristics of each cluster, and label the IPs or contents in this cluster as “normal” or “abnormal” based on the summarized patterns of different types of attacks.
\end{enumerate}

Gaussian mixture model (GMM) is a distribution-based clustering constructed with multiple Gaussian distribution components and a probability density function \cite{gmm}. GMM models data points by utilizing Gaussian distribution models with parameters estimated by the expectation-maximization (EM) algorithms. In GMM, each Gaussian component can be denoted by a multivariate Gaussian distribution \cite{gmm}: 
\begin{equation}G(\mathbf{x} \mid \mu, \Sigma)=\frac{1}{(2 \pi)^{\frac{D}{2}}|\Sigma|^{\frac{1}{2}}} e^{-\frac{1}{2}(x-\mu)^{T} \Sigma^{-1}(x-\mu)}\end{equation}
where $\mathbf{x}$ is the data points, $\mu$ is the mean or the expectation of the Gaussian distribution, $\Sigma$ is the covariance, and $D$ is the dimensionality of the dataset. 

A GMM with $K$ Gaussian components models the data by the following probability density function \cite{gmm}:
\begin{equation}p(\mathbf{x} \mid \theta)=\sum_{i=1}^{K} \pi_{i} G\left(\mathbf{x} \mid \mu_{i}, \Sigma_{i}\right)\end{equation}

where $\theta=\{\pi_{i},\mu_{i},\Sigma_{i}\}$ are the parameters of GMM, $\pi_{i}$ is the weight of each Gaussian component, and $\sum_{i=1}^{K} \pi_{i}=1$.

The GMM parameters are obtained by the EM algorithm that repeats two main steps until convergence: the E-step calculates the expectation of each Gaussian component, and the M-step maximizes the calculated expectations to update the parameters of Gaussian distributions \cite{gmm}. The time complexity of training a GMM is $O(NKD^2)$ for $N$ data instances, $K$ Gaussian components, and $D$ features or dimensions \cite{gmm2}.

The main reasons for choosing GMM for the content-based and IP-based datasets are:
\begin{enumerate}
\item Based on the visualization of the probability density functions, most of the extracted features on the content and client IP sides follow Gaussian or near-Gaussian distributions. Additionally, GMM considers feature covariance and can model more flexible cluster shapes than many other clustering methods, like k-means, which can only return globular cluster shapes. Therefore, GMM can fit the datasets effectively. 
\item Unlike many other clustering algorithms, like k-means and hierarchical clustering, which can only return a cluster label or identity number, GMM is able to give a confidence value to each test sample, indicating the probability of belonging to each cluster. The probability can be used to find uncertain samples and take further actions to reduce errors. 
\item Although k-means has a training time complexity of $O(NKD)$ \cite{kmeans} that is lower than GMM, the training time of GMM is still low because the proposed feature engineering method has effectively reduced the dimensionality of the data. On the other hand, the run-time complexity of a GMM is also $O(NKD)$, so the execution time of running an trained GMM is low. 
\end{enumerate}

For the GMM applied to the CDN datasets, it has a major hyper-parameter that requires tuning, which is $K$, the number of clusters or Gaussian components \cite{hpme}. Identifying an optimal value of $K$ is crucial since it determines whether a sufficient number of Gaussian components are constructed to describe and distinguish normal and abnormal data patterns. On the other hand, a too-large $K$ will lead to additional model training time.

Since GMMs only have a discrete hyper-parameter, the number of Gaussian components, that requires tuning in most cases, BO-GP serves as an effective HPO method for GMMs \cite{hpme}. The silhouette coefficient is also selected as the metric of the GMM model since it measures how similar a data sample is to other data samples within the same cluster and how different a data sample is from the samples in other clusters.

After grouping the contents and IPs into optimized numbers of clusters using GMM and BO-GP, the characteristics of each cluster will be analyzed based on the DoS \& CPA patterns, and the clusters that match the abnormal content and IP patterns summarized in Tables \ref{cpapattern} \& \ref{dospattern} are deemed to have passed the initial detection. The IPs and contents in these clusters are preliminarily labeled “abnormal” at this stage.

\subsection{Multi-perspective Result Validation}
Using unsupervised machine learning algorithms, including GMM and iForest, enables us to distinguish numerically abnormal content, nodes, and IPs from normal ones. However, certain legitimate events, like misconfigurations and crowd events, may perform similar behaviors as cyber-attacks and be misclassified as anomalies. Therefore, a multi-perspective result validation analysis is performed to eliminate the false alarms and improve the detection rate, so as to identify the real network entities affected by CPAs and DoS attacks.
\subsubsection{Time-series Analysis}
At the first stage of result validation, time series analysis is conducted by analyzing the changes of certain features in periods (\textit{e.g.}, hourly and daily changes) to find the abnormal events and potential attacks. 

The major feature changes to be monitored in general CDN access log data for DoS \& CPA detection include follows:
\begin{enumerate}
\item The changes in the hourly number of requests can help us to find potential crowd events and DoS attacks when there is a sudden burst of requests in certain time periods; 
\item The changes in the hourly cache hit rate can help us to detect potential DoS attacks when there is a sudden burst of error requests that aim to exhaust network resources; 
\item The changes in the hourly request popularity can be used to identify potential CPAs when certain IPs start to send a large number of requests for unpopular contents, or certain nodes get many requests for unpopular contents. 
\end{enumerate}

\begin{figure}
     \centering
     \includegraphics[width=6.5cm]{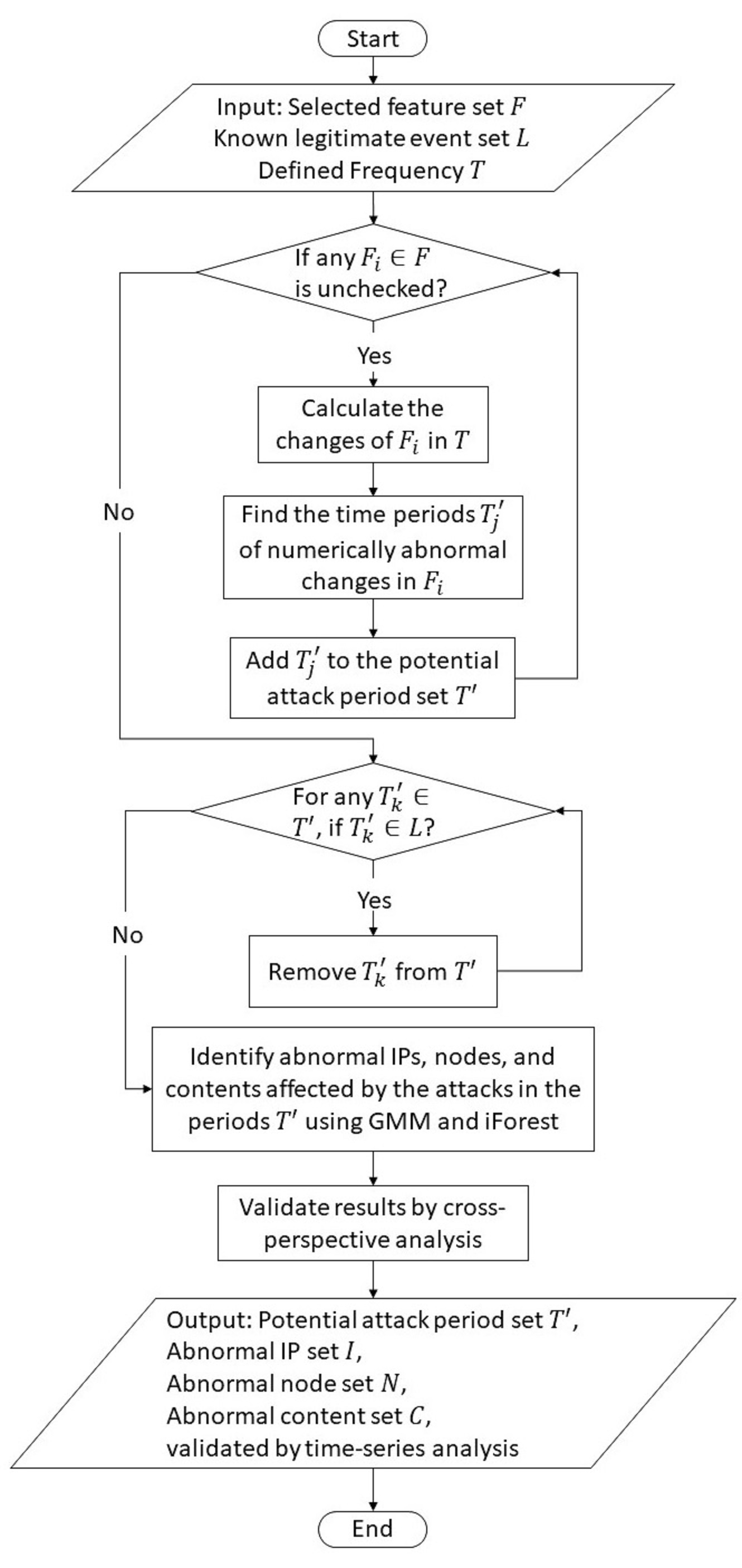}
     \caption{The flow chart of time-series analysis} \label{timeflow}
\end{figure}

The general process of time-series validation is shown in Fig. \ref{timeflow}. After we find the time periods in which certain features change abnormally, they will be compared with the information about the known legitimate events provided by the CDN operator. If known legitimate events did not occur in these abnormal periods, there is a high probability that cyber-attack occurred. Thus, the active IPs, nodes, and contents in these abnormal periods will be analyzed using the proposed optimized iForest or GMM methods to detect the abnormal network entities affected in potential cyber-attacks. 
In most real-world applications, legitimate event information should be recorded for network maintenance purposes. In case of no information about legitimate events, expert intervention can be involved in the validation process as a HITL procedure to help determine the real attacks. Moreover, network administrators can still use authentication mechanisms to confirm the identities of all these suspicious entities to identify real attacks.

\begin{figure}
     \centering
     \includegraphics[width=6.3cm]{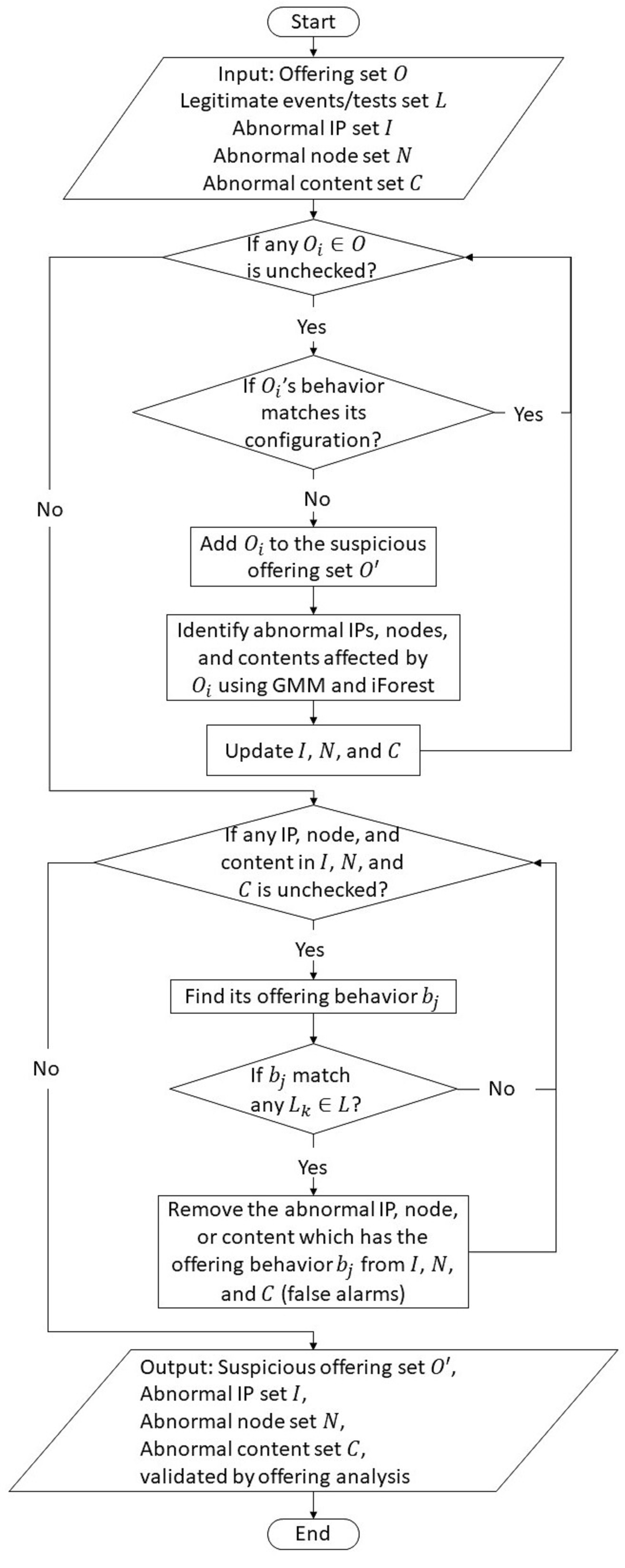}
     \caption{The flow chart of account offering analysis} \label{aoflow}
\end{figure}

The cross-perspective analysis is also utilized in the time-series validation process to help validate the real abnormal network entities. To perform the cross-perspective analysis, the results from each perspective are used to validate the results from the other two perspectives. To be specific, the information about the detected compromised nodes can help us to validate the potential malicious IPs which try to attack these nodes and the potential abnormal contents which are used to pollute these nodes. For the abnormal IPs and contents detected from their own perspectives, these results can be used to validate which nodes are targeted by the attackers who utilized these abnormal IPs and contents. Through this process, the true abnormal entities that are affected by attacks can be identified effectively.

In conclusion, time series analysis enables us to locate the specific days or time periods of potential attacks and legitimate events (\textit{e.g.}, crowd events), thus validating the results by analyzing the nodes, IPs, and contents affected during these periods.

\subsubsection{Account-Offering Analysis}
The account-offering (AO) analysis is to analyze anomaly detection results based on the AO configurations and behaviors to distinguish true attacks from legitimate outliers. The potential behavior of an AO can often be estimated based on its configuration. If cyber-attacks occur, the characteristics of the affected AOs can change to abnormal, which can help us to validate the abnormal network entities serviced by these AOs. For example, if an AO that is configured to be the major stream of static content services gets numerous requests for non-existent live streaming contents, DoS attacks might be launched through this AO to overwhelm certain nodes. On the other hand, if an AO is used for the legitimate tests of old progressive download videos, it may be misclassified as anomalies by the proposed unsupervised models because, similar to CPAs, there will also be a large number of requests for low-popularity contents; thus, the affected nodes, IPs, and contents can be false alarms. AO analysis requires the configuration information of AOs to determine whether the AOs' behaviors match their configurations. The general process of account-offering analysis is shown in Fig. \ref{aoflow}. Through this process, the false alarms from legitimate events will be removed to improve the anomaly detection accuracy.

After implementing the optimized unsupervised anomaly detection method and performing the multi-perspective result analysis, the abnormal contents, compromised nodes, and malicious IPs with their potential attack types can be identified effectively. 

\section{Experimental Results \& Discussion}
\subsection{Experiments Setup}
The experiments are conducted on a machine with a 6 Core i7-8700 processor and 16 GB of memory. The dataset used for experiments is 169 GB of 8 days web access logs collected by a major CDN operator from December 12th to 19th, 2016, including 452,264,816 unique requests/samples. Through the proposed multi-perspective feature engineering, we have obtained the IP-based dataset (1,268,027 unique IPs) and the node-based dataset (50 unique nodes) for abnormal IP and node detection. Additionally, the content-based dataset (1,867,584 unique contents) and the account-offering dataset (70 unique AOs) are also generated to support anomaly detection result validation. On the other hand, the dates of crowd events are provided by the CDN operator, including 12th, 14th, and 15th, December 2016 (days 1, 3, and 4). 
Additionally, the CDN operator has used many conventional security mechanisms as the first layer of defense, including firewalls with filtering mechanisms, authentication, hashing, and load balancers, to mitigate cyber-attacks. Our proposed anomaly detection system serves as the second layer of defense to detect anomalies that are not stopped by the first layer of defense.

At the first stage of the experiments, the anomaly detection model based on iForest and GMM is implemented on the obtained content-based, client IP-based, and node-based datasets individually to preliminarily detect abnormal contents, malicious IPs, and compromised nodes. Once the numerically abnormal network entities are separated from normal entities by the proposed optimized unsupervised learning model, a multi-perspective analysis is then performed to validate the results as the second stage of the experiments. The validation process includes the time-series analysis based on the known legitimate events periods, and the account-offering analysis based on the account-offering configuration information and their practical behaviors. As such, false alarms can be reduced, and real abnormal network entities can be identified effectively. 
\subsection{Unsupervised Anomaly Detection Results}
\subsubsection{Compromised Node Detection Results}
Based on the extracted node-based datasets, there are 50 unique nodes that have received requests in the 8 days dataset. Through the optimized iForest method that returns the contamination level of 0.22, as well as the comparison between node behaviors and cyber-attack patterns, 11 nodes that have behaved abnormally at least on one day are preliminarily identified as compromised nodes, including the nodes number (No.) 0, 3, 4, 5, 7, 8, 9, 25, 36, 39, and 47. Among the detected abnormal nodes, nodes No. 0, 5, 7, and 25 had a low cache hit rate (less than 0.5), while other abnormal nodes had a relatively high request error rate. 

The mean value of each feature for normal and abnormal nodes is shown in Table \ref{nodepre}. It is shown that the compromised nodes have lower cache hit rates and data transfer rates while having higher request error rates than the normal nodes, which is due to potential network congestion and unavailability caused by the attacks. Additionally, the cache hit rates of the legitimate IPs serviced by the abnormal nodes are much lower than the IPs serviced by the normal nodes. A multi-perspective validation will be conducted at the next stage to evaluate the detected abnormal nodes and reduce errors.

\begin{table}[]
\caption{Mean Values of Each Feature of Normal and Abnormal Nodes in Preliminary Anomaly Detection Using IForest}
\centering
\setlength\extrarowheight{1pt}
\scalebox{0.85}{
\begin{tabular}{|>{\centering\arraybackslash}p{4em}|>{\centering\arraybackslash}p{3em}|>{\centering\arraybackslash}p{4.5em}|>{\centering\arraybackslash}p{5.5em}|>{\centering\arraybackslash}p{4em}|>{\centering\arraybackslash}p{4.6em}|}
\hline
\textbf{Node Labels} & \textbf{Cache hit rate} & \textbf{Legitimate IP cache hit rate} & \textbf{Data transfer rate (MB/s)} & \textbf{Request error rate} & \textbf{Request popularity} \\ \hline
Normal               & 0.886                   & 0.928                                 & 0.696                              & 0.003                       & 0.925                       \\ \hline
Abnormal             & 0.286                   & 0.299                                 & 0.374                              & 0.052                       & 0.961                       \\ \hline
\end{tabular}
}
\label{nodepre}%
\end{table}

\subsubsection{Abnormal Content Detection Results}
1,867,584 unique contents have been requested in the 8 days dataset. A GMM optimized by BO-GP is trained on the content-based dataset to detect potential abnormal contents that might be used by attackers to launch attacks. As the major hyper-parameter of GMM, the optimal number of Gaussian components is found to be 28, which returns the highest silhouette score of 0.96. As shown in Table \ref{contentpre}, the 169 low-popularity contents in cluster No. 28 have got a large number of requests on a couple of target nodes. Therefore, these unpopular contents might have been requested many times by CPA attackers to occupy the cache space, making the legitimate and popular contents get cache misses. Therefore, the 169 contents in cluster No. 28 are preliminarily identified as potential abnormal contents. The clusters No. 1-27 are classified as normal clusters based on the comparison with cyber-attack patterns, and the behaviors of the clusters No. 3-27 are omitted in Table \ref{contentpre}. 

\begin{table}[]
\caption{Average Feature Values of Content Clusters in Preliminary Anomaly Detection Using GMM}
\centering
\setlength\extrarowheight{1pt}
\scalebox{0.85}{
\begin{tabular}{|>{\centering\arraybackslash}p{3.5em}|>{\centering\arraybackslash}p{5em}|>{\centering\arraybackslash}p{5em}|>{\centering\arraybackslash}p{5em}|>{\centering\arraybackslash}p{4em}|>{\centering\arraybackslash}p{4.6em}|}
\hline
\textbf{Content cluster No.} & \textbf{Avg number of requests} & \textbf{Avg request per node ratio} & \textbf{Avg request per IP ratio} & \textbf{Avg cache hit rate} & \textbf{Avg popularity} \\ \hline
1                    & 1.0                             & 1.0                                 & 1.0                               & 0.134                       & 0.0                     \\ \hline
2                    & 225.7                           & 8.2                                 & 1.2                               & 0.857                       & 1.0                     \\ \hline
…                    & …                               & …                                   & …                                 & …                           & …                       \\ \hline
28                   & 1021.4                          & 601.2                               & 580.8                             & 0.835                       & 0.254                   \\ \hline
\end{tabular}
}
\label{contentpre}%
\end{table}

\subsubsection{Malicious Client IP Detection Results}
There are 1,268,027 unique IP addresses in the extracted IP-based datasets. For abnormal IP detection, two GMMs are trained on two different feature sets separately to detect CPAs and DoS attacks, respectively. 

For DoS attack detection, the major considered features are the number of requests, requests per node ratio, and average request interval. By using BO-GP to optimize the GMM, the optimal number of clusters is found to be 47 with the highest silhouette score (0.55). As shown in Table \ref{ipdospre}, the 310 IPs in cluster No. 47 are detected as potential DoS attack IPs, since they have sent a large number of requests to several target nodes at a very high frequency. Their request popularity is very high (99.8\%), so they are unlikely to be CPA IPs. 

\begin{table}[]
\caption{Average Feature Values of IP Clusters in Preliminary DoS Attack Detection Using GMM}
\centering
\setlength\extrarowheight{1pt}
\scalebox{0.85}{
\begin{tabular}{|>{\centering\arraybackslash}p{2.9em}|>{\centering\arraybackslash}p{3.8em}|>{\centering\arraybackslash}p{3.5em}|>{\centering\arraybackslash}p{3.5em}|>{\centering\arraybackslash}p{2.5em}|>{\centering\arraybackslash}p{3.3em}|>{\centering\arraybackslash}p{4.6em}|}
\hline
\textbf{IP cluster No.} & \textbf{Avg number of requests} & \textbf{Avg number of nodes} & \textbf{Avg request interval (s)} & \textbf{Avg cache hit rate} & \textbf{Avg request error rate} & \textbf{Avg request popularity} \\ \hline
1                    & 4.2                             & 1.422                        & 1.44                              & 0.925                       & 0.0                             & 0.976                           \\ \hline
2                    & 116.4                           & 4.12                         & 365.8                             & 0.908                       & 0.011                           & 0.957                           \\ \hline
…                    & …                               & …                            & …                                 & …                           & …                               & …                               \\ \hline
47                   & 405607.1                        & 15.6                         & 0.673                             & 0.868                       & 0.124                           & 0.998                           \\ \hline
\end{tabular}
}
\label{ipdospre}%
\end{table}

On the other hand, for CPA detection, the major considered features are the number of requests, request per content ratio, and request popularity. 34 clusters are found to be able to effectively distinguish potential CPA IPs, returning an optimized GMM with the silhouette score of 0.47. The behavior of each cluster is compared with the FLA and LDA patterns summarized in Section IV-C. As shown in Table \ref{ipcpapre}, the 21 IPs in cluster No. 33 are detected as the potential FLA IPs since they have sent a large number of requests to a couple of target nodes for several unpopular contents. Similarly, the 33 IPs in cluster No. 34 are detected as the potential LDA IPs since they have sent a large number of total requests to many unpopular contents to occupy the cache space. Thus, 54 unique IPs are preliminarily identified as potential CPA IPs. 

\begin{table}[]
\caption{Average Feature Values of IP Clusters in Preliminary CPA Detection Using GMM}
\centering
\setlength\extrarowheight{1pt}
\scalebox{0.85}{
\begin{tabular}{|>{\centering\arraybackslash}p{2.9em}|>{\centering\arraybackslash}p{3.8em}|>{\centering\arraybackslash}p{3.5em}|>{\centering\arraybackslash}p{5.0em}|>{\centering\arraybackslash}p{2.5em}|>{\centering\arraybackslash}p{3.3em}|>{\centering\arraybackslash}p{4.6em}|}
\hline
\textbf{IP cluster No.} & \textbf{Avg number of requests} & \textbf{Avg number of nodes} & \textbf{Avg request per content ratio} & \textbf{Avg cache hit rate} & \textbf{Avg request error rate} & \textbf{Avg request popularity} \\ \hline
1                    & 1.3                             & 1.14                         & 1.0                                                                                      & 0.057                       & 0.019                           & 0.500                           \\ \hline
2                    & 200.4                           & 28.4                         & 1.0                                                                                      & 0.699                       & 0.007                           & 0.627                           \\ \hline
…                    & …                               & …                            & …                                                                                        & …                           & …                               & …                               \\ \hline
33                   & 4426.6                          & 1.90                         & 769.8                                                                                    & 0.856                       & 0.040                           & 0.186                           \\ \hline
34                   & 16028.4                         & 5.18                         & 1.1                                                                                      & 0.077                       & 0.019                           & 0.359                           \\ \hline
\end{tabular}
}
\label{ipcpapre}%
\end{table}
\subsection{Multi-perspective Results Validation}
Although ML algorithms can identify the numerically abnormal IPs and nodes, many false alarms may occur, and some real abnormal IPs and nodes may also be ignored. Thus, multi-perspective validation is done based on the in-depth analysis of the datasets, including time series analysis and account-offering analysis. 
\subsubsection{Time Series Analysis}
\begin{figure}
     \centering
     \includegraphics[width=8cm]{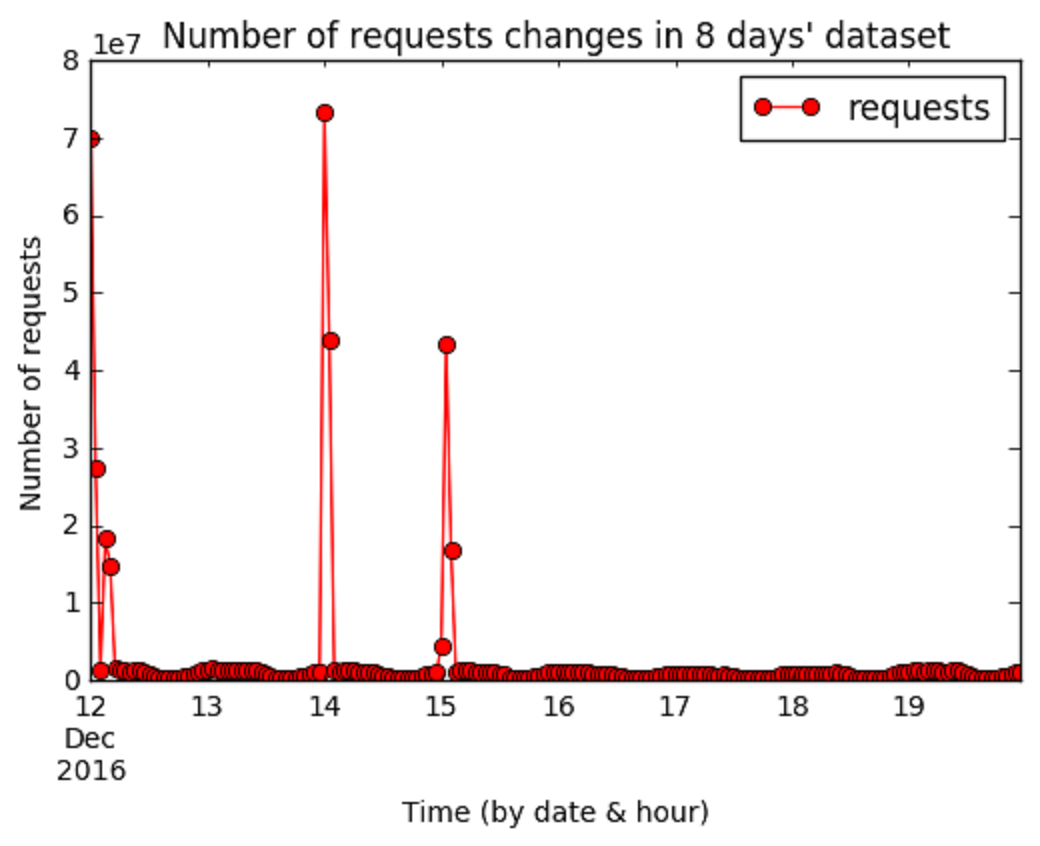}
     \caption{The hourly number of requests change in 8 days dataset.} \label{timerequest}
\end{figure}

Since only the date information about the crowd events is provided by the CDN operator, the hourly number of requests is calculated to find the specific crowd event time periods, as shown in Fig. \ref{timerequest}. According to the hourly number of requests changes, the legitimate event time periods are found to be 00:00-05:00 on day 1, 00:00-02:00 on day 3, and 00:00-03:00 on day 4. By extracting and analyzing the IP-based dataset in these time periods, 250 IPs that have only sent requests in the crowd events have been detected. The daily behavior of these 250 IPs is shown in Table \ref{crowdIP}. Although they have sent a large number of requests at a very high frequency, since they have only sent requests in the crowd events, and other legitimate IPs in the crowd events got more than 99\% cache hit rates, these 250 IPs should belong to legitimate IPs. After checking the 310 IPs that are preliminarily detected as DoS attack IPs by the GMM and described in the last subsection, these 250 legitimate event IPs are removed from the abnormal IP list and labeled normal. Thus, 250 false alarms are removed, and the number of potential DoS attack IPs is reduced to 60 after the first stage of time-series analysis. 

\begin{table}[]
\caption{Main Feature Values of 250 Crowd Event IPs}
\centering
\setlength\extrarowheight{1pt}
\scalebox{0.85}{
\begin{tabular}{|>{\centering\arraybackslash}p{2em}|>{\centering\arraybackslash}p{5em}|>{\centering\arraybackslash}p{4em}|>{\centering\arraybackslash}p{5em}|>{\centering\arraybackslash}p{4em}|>{\centering\arraybackslash}p{4.6em}|}
\hline
\textbf{Day} & \textbf{Avg number of requests} & \textbf{Avg number of nodes} & \textbf{Avg request interval (s)} & \textbf{Avg cache hit rate} & \textbf{Avg request popularity} \\ \hline
1                                  & 501600.6                                             & 21.5                                              & 0.021                                                  & 0.997                                            & 1.0                                                  \\ \hline
3                                  & 459155.2                                             & 10.5                                              & 0.061                                                  & 0.996                                            & 1.0                                                  \\ \hline
4                                  & 491830.9                                             & 26.0                                              & 0.014                                                  & 0.997                                            & 1.0                                                  \\ \hline
\end{tabular}
}
\label{crowdIP}%
\end{table}

\begin{figure}
     \centering
     \includegraphics[width=8.2cm]{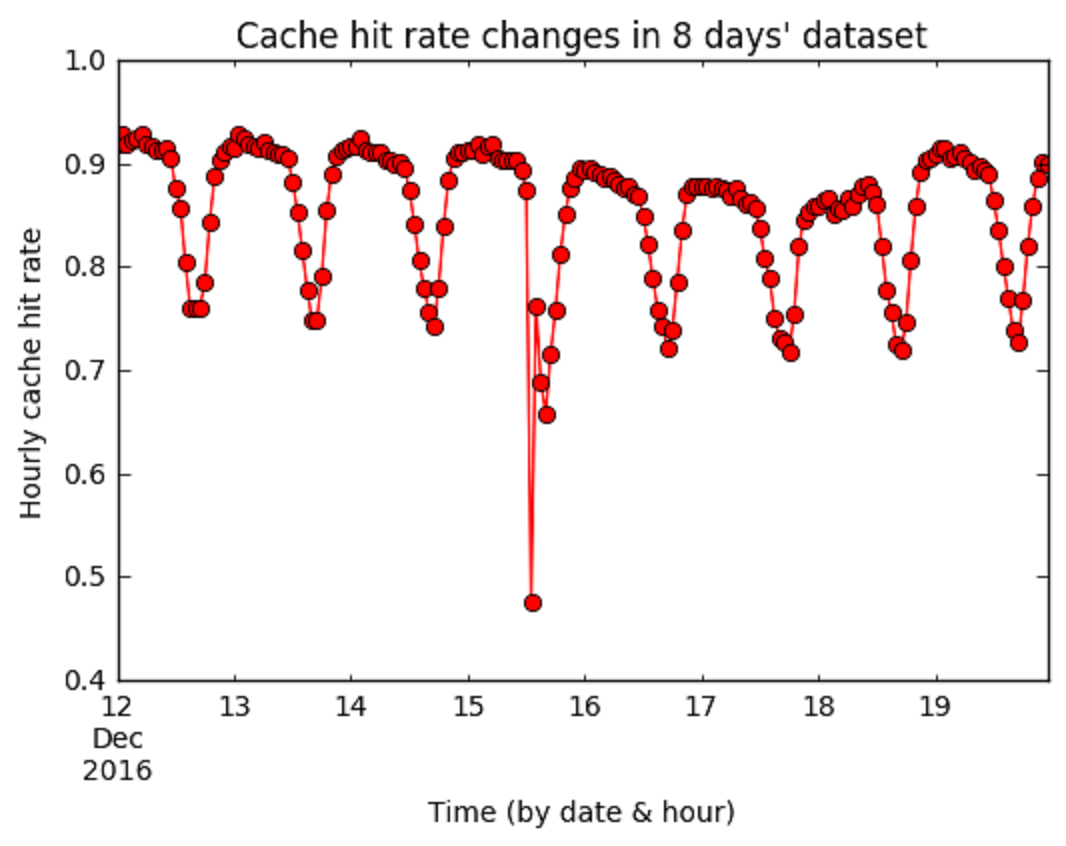}
     \caption{The hourly cache hit rate change in 8 days dataset. } \label{timecpa}
\end{figure}

At the next stage, in order to find the potential cyber-attack time periods, the hourly cache hit rate changes are calculated since the reduced cache hit rate is the most representative behavior of a CDN node that is under a CPA or DoS attack. As shown in Fig. \ref{timecpa}, the hourly cache hit rate remains stationary during the eight days except for 13:00-15:00, day 4. An abnormal event might occur in this time period, so this time period is labeled as an abnormal event period. 

To determine the root cause of the significant cache hit rate drop, we have analyzed the requests, nodes, and client IPs in this abnormal period. Firstly, most of the requests in this abnormal period have got the status code 404, indicating resource-not-found errors. Thus, attackers have sent many requests for non-existing contents to cause many cache misses. A large number of 404-error requests will exhaust the network resources and cause network unavailability or congestion, so a DoS attack might occur during this abnormal period.

In the next step, we have explored the compromised nodes in this abnormal period. It is found that most of the 404-error requests have been sent to node No. 25, so it is the major target of the attack. Additionally, 10 other nodes are found to be the secondary targets since they also got many 404-error requests during the abnormal period, but not as many as node No. 25. As shown in Table \ref{ddosnode}, node No. 25 had a very low cache hit rate (0.21) and a very high request error rate (0.61) on day 4. The other 10 compromised nodes’ cache hit rates are also slightly reduced on day 4, and their request error rates on day 4 are much higher than their error rates on other days. By comparing the compromised nodes in this abnormal event period with the abnormal nodes detected by iForest in Section V-B, three more abnormal nodes missed by the iForest but detected by the time-series validation process, nodes No. 28, 33, and 42, are added to the detected abnormal node list. Although these three nodes’ cache hit rates are not very low (more than 76\%), they have also been affected by the potential cyber-attacks on day 4. Thus, 14 unique nodes are labeled as abnormal nodes at this stage.

\begin{table}[]
\caption{Main Feature Values For Compromised Nodes Under Potential DDoS Attacks}
\centering
\setlength\extrarowheight{1pt}
\scalebox{0.85}{
\begin{tabular}{|>{\centering\arraybackslash}p{2em}|>{\centering\arraybackslash}p{4em}|>{\centering\arraybackslash}p{5em}|>{\centering\arraybackslash}p{4.8em}|>{\centering\arraybackslash}p{5.8em}|>{\centering\arraybackslash}p{5.5em}|}
\hline
\textbf{Node No.} & \textbf{Cache hit rate on day 4} & \textbf{Avg cache hit rate on    other days} & \textbf{Request error rate on day 4} & \textbf{Avg request error rate    on other days} & \textbf{Number of requests on    day 4} \\ \hline
3                 & 0.65                             & 0.70                                         & 0.11                                    & 0.0009                                           & 13,922                                  \\ \hline
4                 & 0.64                             & 0.74                                         & 0.02                                    & 0.0006                                           & 50,947                                  \\ \hline
8                 & 0.62                             & 0.78                                         & 0.10                                    & 0.0008                                           & 16,054                                  \\ \hline
9                 & 0.73                             & 0.79                                         & 0.03                                    & 0.0009                                           & 15,057                                  \\ \hline
25                & 0.21                             & -                                            & 0.61                                    & -                                                & 531,044                                 \\ \hline
28                & 0.76                             & 0.89                                         & 0.08                                    & 0.0014                                           & 52,690                                  \\ \hline
33                & 0.78                             & 0.86                                         & 0.04                                    & 0.0013                                           & 59,415                                  \\ \hline
36                & 0.54                             & 0.75                                         & 0.13                                    & 0.0014                                           & 10,873                                  \\ \hline
39                & 0.70                             & 0.75                                         & 0.04                                    & 0.0016                                           & 11,893                                  \\ \hline
42                & 0.76                             & 0.87                                         & 0.09                                    & 0.0023                                           & 46,475                                  \\ \hline
47                & 0.62                             & 0.85                                         & 0.19                                    & 0.0023                                           & 93,482                                  \\ \hline
\end{tabular}
}
\label{ddosnode}%
\end{table}

\begin{table}[]
\caption{Average Feature Values of IP Clusters in DDoS Attack Detection using GMM}
\centering
\setlength\extrarowheight{1pt}
\scalebox{0.85}{
\begin{tabular}{|>{\centering\arraybackslash}p{2.8em}|>{\centering\arraybackslash}p{3.8em}|>{\centering\arraybackslash}p{3.5em}|>{\centering\arraybackslash}p{3.5em}|>{\centering\arraybackslash}p{2.5em}|>{\centering\arraybackslash}p{3.3em}|>{\centering\arraybackslash}p{4.6em}|}
\hline
\textbf{Cluster No.} & \textbf{Avg number of requests} & \textbf{Avg number of nodes} & \textbf{Avg request interval (s)} & \textbf{Avg cache hit rate} & \textbf{Avg request error rate} & \textbf{Avg request popularity} \\ \hline
1                    & 1.09                            & 1.08                         & 24.9                              & 0.850                       & 0.0                             & 0.941                           \\ \hline
2                    & 354.6                           & 3.78                         & 6.76                              & 0.579                       & 0.0001                          & 0.810                           \\ \hline
…                    & …                               & …                            & …                                 & …                           & …                               & …                               \\ \hline
10                   & 3039.9                          & 2.39                         & 0.6                               & 0.026                       & 0.972                           & 0.947                           \\ \hline
\end{tabular}
}
\label{ddosip}%
\end{table}

After that, the 11,834 client IPs that have sent requests in the abnormal event period are also analyzed. An optimized GMM is utilized to cluster these 11,834 IPs into 10 clusters, which has the highest silhouette score of 0.73. As shown in Table \ref{ddosip}, among the 10 clusters, the 103 IPs in the cluster No. 10 are found to be the malicious IPs since they have sent a large number of 404-error requests to an average of 2.39 target nodes at a high frequency. Thus, these 103 IPs are likely to have launched a distributed DoS (DDoS) attack in the abnormal period 13:00-15:00, day 4. By comparing these 103 IPs with the 60 potential DoS attack IPs detected by the optimized GMM and crowd event analysis, 41 IPs are the overlaps, and the total number of DoS attack IPs is increased to 122. 

\begin{table}[]
\caption{The Behaviors of Suspicious Account-offerings}
\centering
\setlength\extrarowheight{1pt}
\scalebox{0.85}{
\begin{tabular}{|>{\centering\arraybackslash}p{4.4em}|>{\centering\arraybackslash}p{4.8em}|>{\centering\arraybackslash}p{4em}|>{\centering\arraybackslash}p{4em}|>{\centering\arraybackslash}p{5em}|>{\centering\arraybackslash}p{4.6em}|}
\hline
\textbf{AO No.}  & \textbf{Number of requests} & \textbf{Number of nodes} & \textbf{Cache hit rate} & \textbf{Service type} & \textbf{Request popularity} \\ \hline
1                & 44,985,524                  & 27                       & 0.999                   & Static                & 1.0                         \\ \hline
2                & 78,505,897                  & 28                       & 0.999                   & Static                & 1.0                         \\ \hline
3                & 57,276,415                  & 34                       & 0.997                   & Static                & 1.0                         \\ \hline
4                & 63,844,994                  & 25                       & 0.997                   & Static                & 1.0                         \\ \hline
5                & 184,294                     & 34                       & 0.809                   & Progressive download  & 0.262                       \\ \hline
6                & 1,487,0894                  & 47                       & 0.747                   & Live streaming        & 0.762                       \\ \hline
7                & 7,557,320                   & 48                       & 0.763                   & Live streaming        & 0.751                       \\ \hline
8                & 127,665,987                 & 49                       & 0.907                   & Static                & 0.985                       \\ \hline
Other static AOs & 6,501,377                   & 5.06                     & 0.992                   & Static                & 0.999                       \\ \hline
\end{tabular}
}
\label{aos}%
\end{table}

Lastly, it can be seen that this DDoS attack cannot be found by only considering the number of requests shown in Fig. \ref{timerequest}. This is because this DDoS attack mainly targeted node No. 25. Although node No. 25 received a large number of malicious requests in this DDoS attack period, this number of requests is still small compared to the total number of requests received by all nodes or in the crowd events. This emphasizes the reasons for monitoring the changes in different features instead of only one feature for time-series analysis; otherwise, real anomalies may be missed. 

\subsubsection{Account-offering (AO) Analysis}
In the final stage of result validation, account-offering analysis is performed on the current anomaly detection results to remove false alarms. 

There are 70 unique AOs in the 8 days dataset. To analyze each AO, its number of requests, number of nodes, cache hit rate, service type, and request popularity are calculated and analyzed together with their configuration information provided by the CDN operator. The information about the suspicious AOs is shown in Table \ref{aos}. Firstly, the AOs No. 1-4 are the major AOs that provide services during crowd events since they have received a large number of requests for high-popularity contents with more than 99\% cache hit rate during the crowd event periods. The 250 detected crowd event IPs mainly used these 4 AOs, which further proves that they are the legitimate IPs instead of DoS attack IPs. Thus, these 250 IPs are the false positives of the proposed unsupervised model.

\begin{table}[]
\caption{Mean Values of Each Feature on All Days for Three Abnormal Nodes}
\centering
\setlength\extrarowheight{1pt}
\scalebox{0.85}{
\begin{tabular}{|>{\centering\arraybackslash}p{1.8em}|>{\centering\arraybackslash}p{2.8em}|>{\centering\arraybackslash}p{4.3em}|>{\centering\arraybackslash}p{4em}|>{\centering\arraybackslash}p{3.3em}|>{\centering\arraybackslash}p{3.6em}|>{\centering\arraybackslash}p{4.6em}|}
\hline
\textbf{Node No.} & \textbf{Avg cache hit rate} & \textbf{Avg legitimate IP cache hit rate} & \textbf{Avg data transfer rate (MB/s)} & \textbf{Avg request error rate} & \textbf{Avg number of requests} & \textbf{Avg request popularity}\\ \hline
0                 & 0.49                        & 0.0                                       & 0.69                                      & 0.0                             & 344             & 0.935                \\ \hline
5                 & 0.06                        & 0.08                                      & 0.13                                      & 0.0                             & 2383.8          & 0.905                \\ \hline
7                 & 0.0002                      & 0.0                                       & 0.0009                                    & 0.0002                          & 531.6          & 0.881                 \\ \hline
\end{tabular}
}
\label{3nodes}%
\end{table}

\begin{table}[]
\caption{Mean Values of Each Feature For Normal and Detected Abnormal Nodes}
\centering
\setlength\extrarowheight{1pt}
\scalebox{0.85}{
\begin{tabular}{|>{\centering\arraybackslash}p{4em}|>{\centering\arraybackslash}p{3em}|>{\centering\arraybackslash}p{4.5em}|>{\centering\arraybackslash}p{5.2em}|>{\centering\arraybackslash}p{4em}|>{\centering\arraybackslash}p{4.6em}|}
\hline
\textbf{Node label} & \textbf{Cache hit rate} & \textbf{Legitimate IP cache hit rate} & \textbf{Data transfer rate    (MB/s)} & \textbf{Request error rate} & \textbf{Request popularity} \\ \hline
Normal               & 0.936                   & 0.928                                 & 0.805                                 & 0.004                       & 0.978                       \\ \hline
Abnormal             & 0.405                   & 0.416                                 & 0.506                                 & 0.049                       & 0.936                       \\ \hline
\end{tabular}
}
\label{nodefinal}%
\end{table}

For the detected abnormal IPs and nodes, the AOs No. 5-8 shown in Table \ref{aos} are the four suspicious AOs since almost all the requests of the detected potential abnormal IPs and nodes were sent through these 4 AOs. It is found in the configuration information that the node No. 5 is the major progressive download content source, and certain unpopular contents might have been requested by an IP many times for being progressively downloaded. The AOs No. 6 \& 7 are the major live streaming content sources and have relatively high cache hit rates and request popularity, about 75\%. Thus, certain IPs that have used any of these two AOs but got a very low cache hit rate or request popularity are likely to be abnormal. Moreover, the AO No. 8 is the major static content source. It can be seen in the last row of Table \ref{aos} that other static AOs have got a more than 99\% cache hit rate and request popularity, much higher than the AO No. 8, so the AO No. 8 might be used by certain IPs to launch attacks.

Further result validation is done through the AO analysis. Firstly, among the 14 detected abnormal nodes, the 11 compromised nodes under potential DDoS attacks listed in Table \ref{ddosnode} are validated to be the real compromised nodes since their major AOs belong to the abnormal AOs No. 6-8. For the other three detected abnormal nodes (nodes No. 0, 5, 7) shown in Table \ref{3nodes}, since their AOs are legitimate AOs and they have not received many requests (less than 2,500), so they might have been breached before day 1, so any new legitimate requests were unable to get cache hits. Additionally, nodes No. 5 \& 7 were attacked by CPAs since they have much lower request popularities than normal nodes (0.905 \& 0.881 versus 0.978). Therefore, a total of 14 compromised nodes, including 3 nodes that have already been attacked before day 1, and 11 nodes that were under attack on day 4, have been identified and illustrated in Tables \ref{ddosnode} and \ref{3nodes}, and their average feature values are shown in Table \ref{nodefinal}.

\begin{table}[]
\caption{Mean Values of Each Feature for Normal and Detected Abnormal Contents}
\centering
\setlength\extrarowheight{1pt}
\scalebox{0.85}{
\begin{tabular}{|>{\centering\arraybackslash}p{3.5em}|>{\centering\arraybackslash}p{4em}|>{\centering\arraybackslash}p{4.5em}|>{\centering\arraybackslash}p{4em}|>{\centering\arraybackslash}p{4em}|>{\centering\arraybackslash}p{4.6em}|}
\hline
\textbf{Content label} & \textbf{Avg number of requests} & \textbf{Avg request per node ratio} & \textbf{Avg request per IP ratio} & \textbf{Avg cache hit rate} & \textbf{Avg popularity} \\ \hline
Normal                  & 95.6                            & 7.7                                 & 1.5                               & 0.21                        & 0.19                    \\ \hline
Abnormal                & 1049.7                          & 228.9                               & 284.3                             & 0.852                       & 0.333                   \\ \hline
\end{tabular}
}
\label{contentfinal}%
\end{table}

\begin{table}[]
\caption{Mean Values of Each Feature For Normal and Detected Potential CPA IPs}
\centering
\setlength\extrarowheight{1pt}
\scalebox{0.85}{
\begin{tabular}{|>{\centering\arraybackslash}p{2.8em}|>{\centering\arraybackslash}p{3.8em}|>{\centering\arraybackslash}p{3.5em}|>{\centering\arraybackslash}p{5.0em}|>{\centering\arraybackslash}p{2.5em}|>{\centering\arraybackslash}p{3.3em}|>{\centering\arraybackslash}p{4.6em}|}
\hline
\textbf{IP label} & \textbf{Avg number of requests} & \textbf{Avg number of nodes} & \textbf{Avg request per content ratio} & \textbf{Avg cache hit rate} & \textbf{Avg request error rate} & \textbf{Avg request popularity} \\ \hline
Normal             & 161.3                           & 2.1                          & 1.56                                   & 0.925                       & 0.006                           & 0.977                           \\ \hline
LDA                & 9461.4                          & 3.76                         & 1.13                                   & 0.111                       & 0.0006                          & 0.316                           \\ \hline
FLA                & 9418.8                         & 5.5                          & 84.74                                  & 0.691                       & 0.006                           & 0.478                           \\ \hline
CPA                & 9458.8                          & 3.87                         & 6.36                                   & 0.147                       & 0.001                           & 0.326                           \\ \hline
\end{tabular}
}
\label{cpaipfinal}%
\end{table}

According to the configuration information provided by the CDN operator, it is known that as a major source for progressive download videos, the AO No. 5 was used for the tests of old videos, most of which have low popularity. Thus, AO No. 5 has some similar characteristics to CPAs since it is also used to request for low popularity files many times, but it was used by legitimate users to do tests. Therefore, for the detected abnormal contents and CPA IPs, those using the AO No.5 as the major AO are found to be legitimate entities and removed from the abnormal content and IP list. After removing the false positives (113 contents and 21 IPs), the behaviors of the detected 56 real abnormal contents and 33 real CPA IPs are shown in Tables \ref{contentfinal} and \ref{cpaipfinal}, respectively.  For the two types of CPAs, 30 IPs are labeled as LDA IPs, and 3 other IPs are labeled as FLA IPs.

Lastly, through AO analysis, all the detected DoS IPs have used the AOs No. 6-8 to make requests, and they are labeled real DoS IPs. Among these 122 DoS IPs, 103 of them are the DDoS attack IPs that have launched a DDoS attack together on day 4, from 13:00 to 15:00. The other 19 abnormal IPs may have tried to launch a DoS attack individually during different time periods. Their behaviors are shown in Table \ref{dosipfinal}.

\begin{table}[]
\caption{Mean Values of Each Feature For Normal and Detected Potential DoS Attack IPs}
\centering
\setlength\extrarowheight{1pt}
\scalebox{0.85}{
\begin{tabular}{|>{\centering\arraybackslash}p{2.8em}|>{\centering\arraybackslash}p{3.8em}|>{\centering\arraybackslash}p{3.5em}|>{\centering\arraybackslash}p{3.5em}|>{\centering\arraybackslash}p{2.5em}|>{\centering\arraybackslash}p{3.3em}|>{\centering\arraybackslash}p{4.6em}|}
\hline
\textbf{IP label} & \textbf{Avg number of requests} & \textbf{Avg number of nodes} & \textbf{Avg request interval (s)} & \textbf{Avg cache hit rate} & \textbf{Avg request error rate} & \textbf{Avg request popularity} \\ \hline
Normal             & 161.3                           & 2.1                          & 1177.6                            & 0.925                       & 0.006                           & 0.977                           \\ \hline
DDoS               & 3039.9                          & 2.39                         & 0.6                               & 0.026                       & 0.972                           & 0.947                           \\ \hline
Other DoS          & 59789.5                         & 4.63                         & 1.97                              & 0.708                       & 0.264                           & 0.983                           \\ \hline
All DoS            & 14465.3                         & 2.98                         & 1.44                              & 0.16                        & 0.821                           & 0.943                           \\ \hline
\end{tabular}
}
\label{dosipfinal}%
\end{table}

\subsection{Results Summary}
In summary, 14 compromised nodes, 56 abnormal contents, 33 CPA IPs, and 122 DoS attack IPs were detected by the proposed anomaly detection model, as shown in Table \ref{summary}. Among the detected abnormal network entities, 12 nodes and 122 IPs were affected by DoS attacks, while 2 nodes, 33 IPs, and 56 contents were affected by CPAs. Additionally, 384 false positives (FPs) and 65 false negatives (FNs) were removed through the multi-perspective validation process.
As the utilized datasets are completely unlabeled, all these anomaly detection results have been analyzed and verified to be 100\% accurate by multiple cybersecurity experts and industrial partner security network engineers. This verification process is a necessary HITL procedure in real-world applications, as the experts and engineers have in-depth knowledge of the dataset, legitimate events, and cyber-attack patterns. This process also validates the effectiveness of the proposed framework. 

Lastly, the performance of the proposed framework is compared with several data analytics and anomaly detection techniques. As discussed in Section IV-D, binary outlier detection algorithms are more suitable for the node-based dataset that only has 50 samples, so two standard binary outlier detection algorithms, iForest \cite{if} and one-class support vector machine (OC-SVM) \cite{ocsvm}, are used for the comparison of abnormal node detection. On the other hand, clustering algorithms are suitable for IP \& content-based datasets that have more than 1 million samples, so two common clustering algorithms, k-means \cite{kmcpa} and GMM \cite{gmm}, are used for the comparison of abnormal IP \& content detection. For the performance metrics, since the datasets are highly imbalanced, precision (Pre), Recall (Rec), and F1-score are used with accuracy (Acc) for model evaluation. By calculating the harmonic mean of the precision and recall, F1-score is a reliable metric to measure the classification performance on imbalanced datasets \cite{treeme}. 

The model performance comparison is shown in Table \ref{comparison}. 
The accuracy of most methods are larger than 99\%, but this is mainly due to the imbalanced dataset (more than 99\% of data samples are normal data). The iForest and GMM models used for the result comparison are only themselves without the proposed validation procedures; hence, many FPs and FNs occurred, which reduced the F1-scores of iForest and GMM to 88.0\%, 35.8\%, and 49.8\% on the node, IP, and content-based datasets, respectively. This emphasizes the importance of implementing the proposed multi-perspective validation method. Moreover, the F1-scores of OC-SVM and k-means algorithms are lower than the iForest and GMM models from all three perspectives, which justifies the rationale for choosing iForest and GMM in our proposed framework. 

\begin{table}[]
\caption{Summary of Detected Anomalies}
\centering
\setlength\extrarowheight{1pt}
\scalebox{0.85}{
\begin{tabular}{|>{\centering\arraybackslash}p{6em}|>{\centering\arraybackslash}p{7.6em}|>{\centering\arraybackslash}p{6.6em}|>{\centering\arraybackslash}p{8.6em}|}
\hline
\textbf{Label} & \textbf{Number of nodes} & \textbf{Number of IPs} & \textbf{Number of contents} \\ \hline
Normal         & 36                       & 1,267,872              & 1,867,528                   \\ \hline
DoS            & 12                       & 122                    & -                           \\ \hline
CPA            & 2                        & 33                     & 56                          \\ \hline
Removed FPs    & 0                        & 271                 & 113                         \\ \hline
Removed FNs    & 3                        & 62                     & 0                           \\ \hline
\end{tabular}
}
\label{summary}%
\end{table}

\begin{table}[]
\caption{Performance Comparison with Regular Anomaly Detection Techniques}
\centering
\setlength\extrarowheight{1pt}
\scalebox{0.85}{
\begin{tabular}{|>{\centering\arraybackslash}p{7em}|>{\centering\arraybackslash}p{5.5em}|>{\centering\arraybackslash}p{3em}|>{\centering\arraybackslash}p{3em}|>{\centering\arraybackslash}p{3em}|>{\centering\arraybackslash}p{3em}|}
\hline
\textbf{Method} & \textbf{Detection perspective} & \textbf{Acc (\%)} & \textbf{Pre (\%)} &\textbf{Rec (\%)} &\textbf{F1 (\%)} \\ \hline
Proposed & All                            & 100.0  & 100.0 & 100.0        & 100.0        \\ \hline
IForest \cite{if}        & \multirow{2}{*}{Node}          & 94.0   &100.0  &  78.6        & 88.0       \\ \cline{1-1} \cline{3-6} 
OC-SVM  \cite{ocsvm}        &                                & 86.0    & 73.3 & 78.6            & 75.9        \\ \hline
GMM \cite{gmm}            & \multirow{2}{*}{IP}            & 99.97     &25.6  &  60.0         & 35.8        \\ \cline{1-1} \cline{3-6} 
K-means  \cite{kmcpa}       &                                & 99.96    & 21.1 &   69.7         & 32.3        \\ \hline
GMM \cite{gmm}             & \multirow{2}{*}{Content}       & 99.99     & 33.1 &   100.0        & 49.8        \\ \cline{1-1} \cline{3-6} 
K-means \cite{kmcpa}        &                                & 99.99    & 25.8 &  100.0          & 41.0        \\ \hline
\end{tabular}
}
\label{comparison}%
\end{table}

\begin{table}[]
\caption{Anomaly Detection Time Comparison}
\centering
\setlength\extrarowheight{1pt}
\scalebox{0.85}{
\begin{tabular}{|>{\centering\arraybackslash}p{4.5em}|>{\centering\arraybackslash}p{6em}|>{\centering\arraybackslash}p{2em}|>{\centering\arraybackslash}p{2em}|>{\centering\arraybackslash}p{2em}|>{\centering\arraybackslash}p{2em}|>{\centering\arraybackslash}p{2em}|>{\centering\arraybackslash}p{2em}|}
\hline
\multirow{3}{*}{\textbf{\begin{tabular}[c]{@{}c@{}}Detection \\ Perspective\end{tabular}}} & \multirow{3}{*}{\textbf{Method}} & \multicolumn{6}{c|}{\textbf{Detection Time of Each Anomaly (s)}}                                                                                                                                                                     \\ \cline{3-8} 
                                                                                           &                                  & \multicolumn{2}{c|}{\textbf{\begin{tabular}[c]{@{}c@{}}Feature \\ Engineering\end{tabular}}} & \multicolumn{2}{c|}{\textbf{\begin{tabular}[c]{@{}c@{}}Unsupervised \\ Detection\end{tabular}}} & \multicolumn{2}{c|}{\textbf{Total}} \\ \cline{3-8} 
                                                                                           &                                  & \textbf{Avg}                                  & \textbf{Max}                                 & \textbf{Avg}                                   & \textbf{Max}                                   & \textbf{Avg}     & \textbf{Max}     \\ \hline
\multirow{2}{*}{Node}                  & Proposed                & \multirow{2}{*}{9.8}    & \multirow{2}{*}{29.6}   & 0.1                      & 0.3                       & 9.9                   & 29.9                  \\ \cline{2-2} \cline{5-8} 
                                       & OC-SVM \cite{ocsvm}          &                         &                         & 0.5                      & 1.6                       & 10.3                  & 31.2                  \\ \hline
\multirow{2}{*}{IP}                    & Proposed                & \multirow{2}{*}{12.4}   & \multirow{2}{*}{32.9}   & 6.8                      & 25.6                      & 19.2                  & 58.5                  \\ \cline{2-2} \cline{5-8} 
                                       & K-means \cite{kmcpa}         &                         &                         & 3.1                      & 15.3                      & 15.5                  & 48.2                  \\ \hline
\multirow{2}{*}{Content}               & Proposed                & \multirow{2}{*}{5.8}    & \multirow{2}{*}{12.5}   & 4.5                      & 10.4                      & 10.3                  & 22.9                  \\ \cline{2-2} \cline{5-8} 
                                       & K-means \cite{kmcpa}         &                         &                         & 2.3                      & 7.7                       & 8.1                   & 20.2                  \\ \hline
\end{tabular}
}
\label{time}%
\end{table}

As the CDN requests/traffic are continuously generated and we aim to detect abnormal network entities (\textit{i.e.}, node, IP, content) and corresponding attacks, the node, IP, and content-based datasets need to be continuously updated by implementing the proposed feature engineering method on the new CDN traffic data. Then, the proposed unsupervised learning models continuously detect anomalies on the updated datasets. These are the major procedures that take time. The reason for the reaction or detection time is because the proposed system needs to process a number of requests to update the behaviors of nodes, IPs, and contents; thus, the anomalies can be identified. To protect CDNs against DoS and CPA, the anomalies should be detected in time. Hence, assuming the traffic is continuously generated, we have evaluated the detection speed of the proposed system by measuring the total execution time from the time each abnormal entity is affected by an attack to the time the proposed method detects this anomaly. This anomaly detection time has been further divided into the feature engineering time and the unsupervised model detection time. Moreover, as different anomalies/attacks show different patterns and need different detection time, we have measured the average (Avg) and maximum (Max) execution time of all abnormal nodes, IPs, and contents in seconds.  

The anomaly detection time of the proposed method and the two compared methods (OC-SVM and k-means) is shown in Table \ref{time}. From the node perspective, the proposed method can detect the abnormal nodes in the average time of 9.9s and the maximum time of 29.9s, which is slighter faster than OC-SVM. The unsupervised detection time for abnormal node detection is low, because the small number of nodes (50) has made it faster for the proposed method to detect the abnormal nodes than the abnormal IPs or contents. From the IP and content perspectives, the proposed method can detect abnormal IPs and contents in the average time of 19.2s and 10.3s, and in the maximum time of 58.5s and 22.9s, respectively. The detection time of the abnormal IPs is higher than the time of the abnormal contents, because detecting certain DoS IPs needs to analyze a large number of requests. Nevertheless, the maximum abnormal IP detection time is still at a low level (58.5s). This shows that the proposed solution can detect anomalies at the early stages of attacks, which can help CDNs stop current attacks in time and prevent future attacks. Although the anomaly detection time of k-means is lower than the proposed method, the proposed method can achieve much higher performance than k-means, as shown in Table \ref{comparison}. Therefore, our proposed method still performs the best among the comparisons by considering both detection accuracy and speed.

\section{Practical Usage and Open Issues}
Given that the proposed framework is working with unlabeled data, the framework can be used as the first level of the anomaly detection process. Unlike traditional anomaly detection processes for unlabeled data, which requires massive manual analysis and expert knowledge, the proposed framework can reduce much overhead using the automatically-tuned ML algorithms and a systematic multi-perspective result validation process proposed in Section IV. Additionally, the high performance of the proposed framework has been verified by multiple security experts.  

One open issue with this work is that it only considers two common service targeting attacks, \textit{i.e.}, CPAs and DoS attacks. Nevertheless, the proposed framework can be extended to new attacks based on the same procedures:
\begin{enumerate}
\item Collect sufficient network log data that contains the new attack samples;
\item Analyze the new attack, summarize the attack patterns, and select appropriate features that can reflect the new attack patterns;
\item Utilize the proposed unsupervised anomaly detection model to detect suspicious activities and compare them with the summarized characteristics of new attacks to determine the real attacks;
\item Utilize the proposed multi-perspective result validation method to remove false alarms and false negatives. 
\end{enumerate}

Through this process, any new attacks that can be reflected from CDN log data can be effectively detected by the proposed method.

On the other hand, although the proposed framework can reduce much manual analysis and labeling process, certain expert knowledge and legitimate event information are still required in the proposed framework to achieve accurate anomaly detection. This is because certain numerical anomalies identified by unsupervised ML algorithms are not true anomalies and require further analysis to distinguish between certain legitimate abnormal events and cyber-attacks. 

Lastly, several procedures in addition to our work should be implemented for real-world CDN applications. 
Since this research work can be considered a pseudo labeling process on an unlabeled raw CDN log dataset, the labeled content, client IP, and node-based datasets can be used to train reliable supervised classifiers. Through this procedure, human efforts will not be required in the future anomaly detection process, and the system can automatically detect anomalies by continuously processing the incoming data.
Thus, supervised model development is one direction for future work.

Moreover, corresponding countermeasures should be made along with attack detection to stop or prevent cyber-attacks. The proposed multi-perspective anomaly detection framework enables the implementation of countermeasures on both attacker and victim sides.

Firstly, on the client or attacker side, traffic filtering and blacklisting are potential response mechanisms to prevent the detected malicious clients from sending requests to CDN servers \cite{counter1}. Detecting abnormal IPs enables network administrators to locate the origins of cyber-attacks \cite{counter2}. Additionally, the domains and specific geographic locations can be found based on the detected malicious client IP addresses. To make countermeasures, the detected malicious IPs, as well as the IPs from the same domains and locations, can be added to the blacklist to block or limit the traffic from these IPs until the identities of these suspicious clients are verified. Hence, the current attacks can be stopped, and future attacks can be prevented by blacklisting the detected malicious IPs \cite{dos_cdn}.  

Secondly, on the victim side, countermeasures can be made from both content and node perspectives. Blacklisting can also be used for detected abnormal contents \cite{counter1}. The contents in the blacklist will be deleted from the cache space of any edge servers. Additionally, the edge servers can reject the requests sent for these abnormal contents or never cache them. Through this process, CPAs can be stopped because the detected abnormal contents cannot be used by CPA attackers to pollute the cache space of CDN servers. On the other hand, the compromised nodes can be isolated until they recover. Specifically, the compromised node information will be notified to other nodes, so that other nodes can avoid communicating with the affected nodes that are under attack \cite{counter1}. The nodes can recover by limiting the requests sent to them or implementing the blacklisting strategies from the client IP and content perspectives. Once recovered, the nodes can be removed from the isolation list and continue communications. 

In conclusion, as the proposed anomaly detection can continuously detect anomalies, any new malicious or compromised network entities can be identified, and corresponding countermeasures can be made to defend against cyber-attacks when they occur. Moreover, the multi-perspective countermeasures enhance the defense capabilities of CDNs. For example, an attacker cannot continue an attack by simply changing IP addresses, because the countermeasures from other perspectives, like isolating nodes and blacklisting contents, can still stop the attack.

Since the development of supervised models and countermeasures is outside the scope of this work, it will be our future work.

\section{Conclusion}
CDNs have become a major content distribution technology in modern networks. However, their caching mechanism introduces additional vulnerabilities. In this paper, we proposed a multi-perspective anomaly detection approach based on a real-world general CDN access log dataset to identify abnormal network entities and cyber-attacks. To detect DoS and cache pollution attacks, we first summarized their patterns and then extracted features from four main perspectives: content, client IP, account-offering, and node perspectives. After obtaining the extracted datasets from multiple perspectives, the anomalies were identified using the optimized unsupervised learning model constructed with the optimized isolation forest and Gaussian mixture models. A comprehensive validation method, including multi-perspective analysis, time-series analysis, and account-offering analysis, was implemented to validate the detected abnormal network entities and the corresponding cyber-attacks. Thus, detection errors can be effectively reduced. Ultimately, the abnormal contents, compromised nodes, and malicious IPs were detected and labeled. In future work, the labeled anomaly detection results can be used for classifier development so that an automated process can be developed to detect new attacks and abnormal network entities effectively. Certain security mechanisms, such as isolating and blacklisting the detected abnormal network entities, can be utilized after anomaly detection to secure CDNs.

\section{Acknowledgment}
This work is partially supported by the Natural Sciences and Engineering Research Council of Canada (NSERC) [NSERC Strategic Partnership Grant STPGP – 521537] and Ericsson Canada.

\ifCLASSOPTIONcaptionsoff
  \newpage
\fi


\begin{IEEEbiography}[{\includegraphics[width=1in,height=1.25in,clip,keepaspectratio]{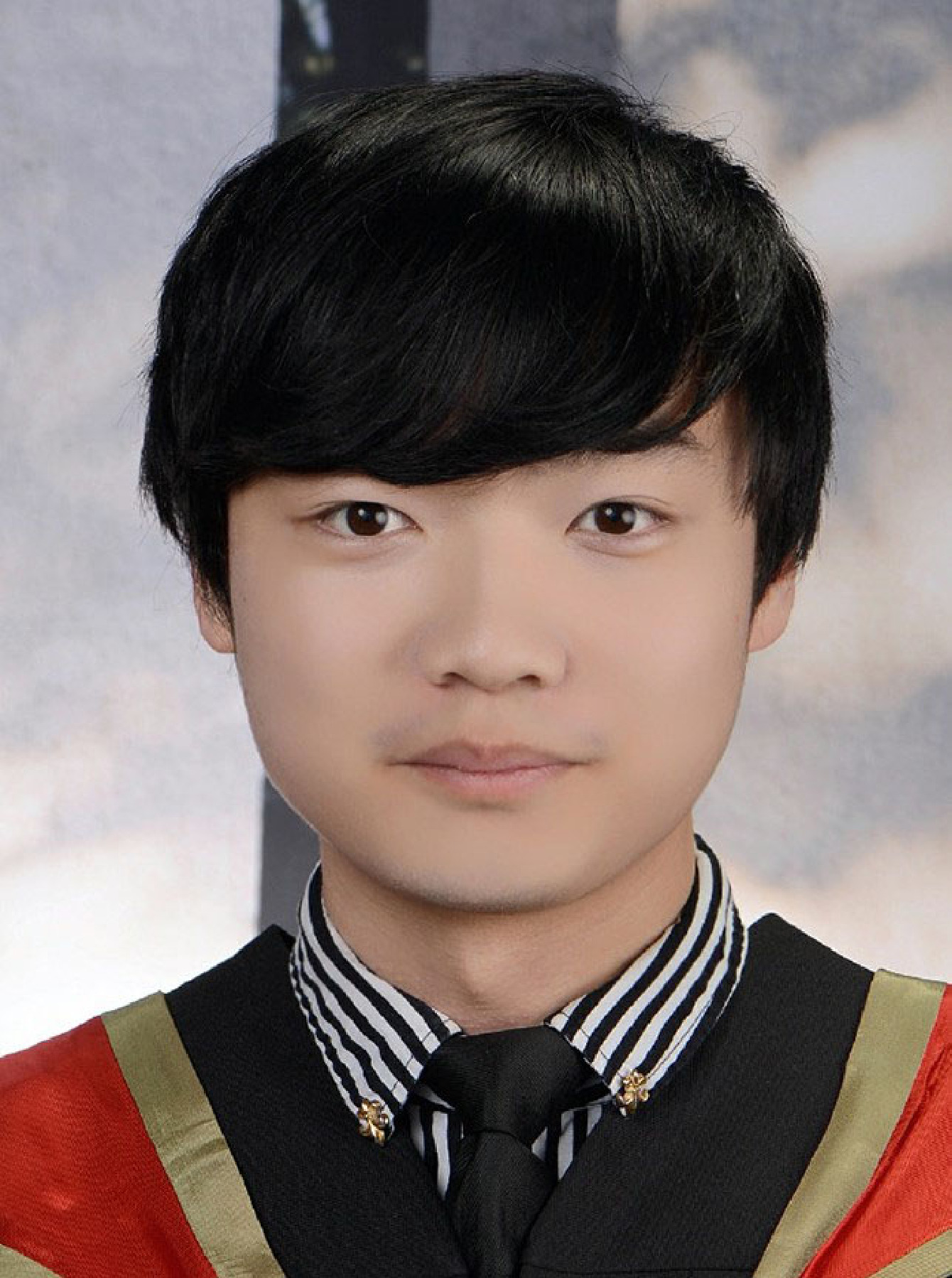}}]{Li Yang}received the B.E. degree in computer science from Wuhan University of Science and Technology, Wuhan, China in 2016 and the MASc degree in Engineering from University of Guelph, Guelph, Canada, 2018. Since 2018 he has been working toward the Ph.D. degree in the Department of Electrical and Computer Engineering, Western University, London, Canada. His research interests include cybersecurity, machine learning, deep learning, and time-series analysis.
\end{IEEEbiography}

\vskip -2pt plus -1fil

\begin{IEEEbiography}[{\includegraphics[width=1in,height=1.25in,clip,keepaspectratio]{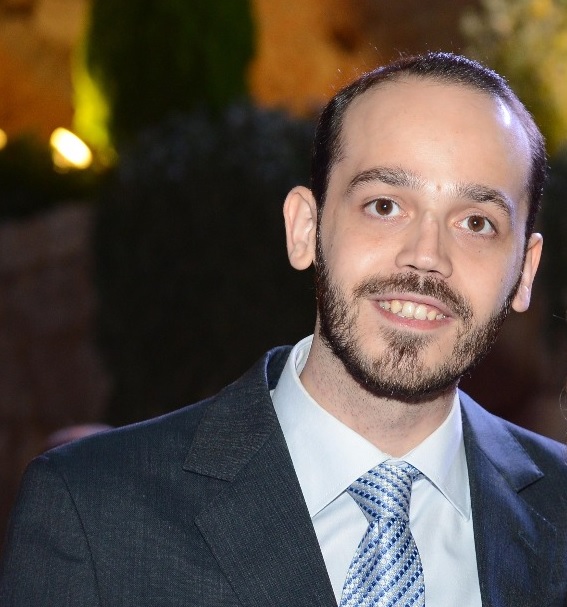}}] {Abdallah Moubayed}received his Ph.D. in Electrical \& Computer Engineering from the University of Western Ontario in August 2018,  his M.Sc. degree in Electrical Engineering from King Abdullah University of Science and Technology, Thuwal, Saudi Arabia in 2014, and his B.E. degree in Electrical Engineering from the Lebanese American University, Beirut, Lebanon in 2012. Currently, he is a Postdoctoral Associate in the Optimized Computing and Communications (OC2) lab at University of Western Ontario.  His research interests include wireless communication, resource allocation, wireless network virtualization, performance \& optimization modeling, machine learning \& data analytics, computer network security, cloud computing, and e-learning.
\end{IEEEbiography}

\vskip -2pt plus -1fil

\begin{IEEEbiography}[{\includegraphics[width=1in,height=1.25in,clip,keepaspectratio]{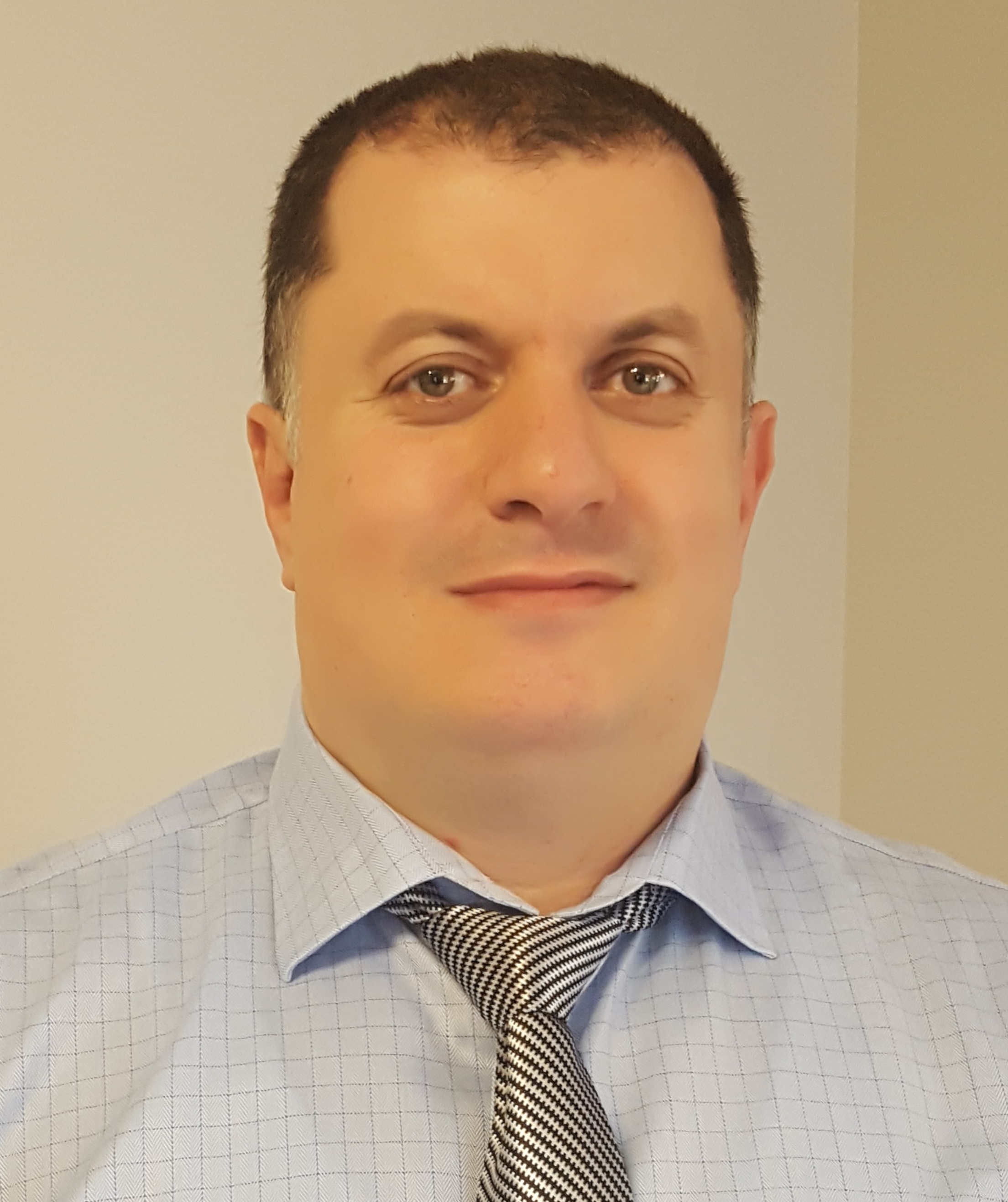}}] {Abdallah Shami}is a professor with the ECE Department at Western University, Ontario, Canada. He is the Director of the Optimized Computing and Communications Laboratory at Western University (https://www.eng.uwo.ca/oc2/). He is currently an associate editor for IEEE Transactions on Mobile Computing, IEEE Network, and IEEE Communications Surveys and Tutorials. He has chaired key symposia for IEEE GLOBECOM, IEEE ICC, IEEE ICNC, and ICCIT. He was the elected Chair of the IEEE Communications Society Technical Committee on Communications Software (2016-2017) and the IEEE London Ontario Section Chair (2016-2018).
\end{IEEEbiography}

\vskip -2pt plus -1fil

\begin{IEEEbiography}[{\includegraphics[width=1in,height=1.25in,clip,keepaspectratio]{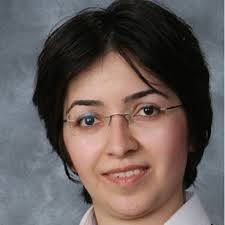}}] {Parisa Heidari}is working as a security master and IoT developer at Ericsson. She received her Masters and PhD in computer engineering from Ecole Polytechnique de Montreal, Canada in 2007 and 2012, respectively. She worked as research associate at Concordia University in collaboration with Ericsson from 2013-2014 and joined Ericsson in 2015.  Her research interests include Internet of Things, Edge Computing, container technologies, function as a Service and different aspects of QoS in cloud system such as optimal resource dimensioning, placement and security of cloud applications. She holds to her credit several patent and publications. She is an expert in the Ericsson cybersecurity team.
\end{IEEEbiography}

\vskip -2pt plus -1fil

\begin{IEEEbiography}[{\includegraphics[width=1in,height=1.25in,clip,keepaspectratio]{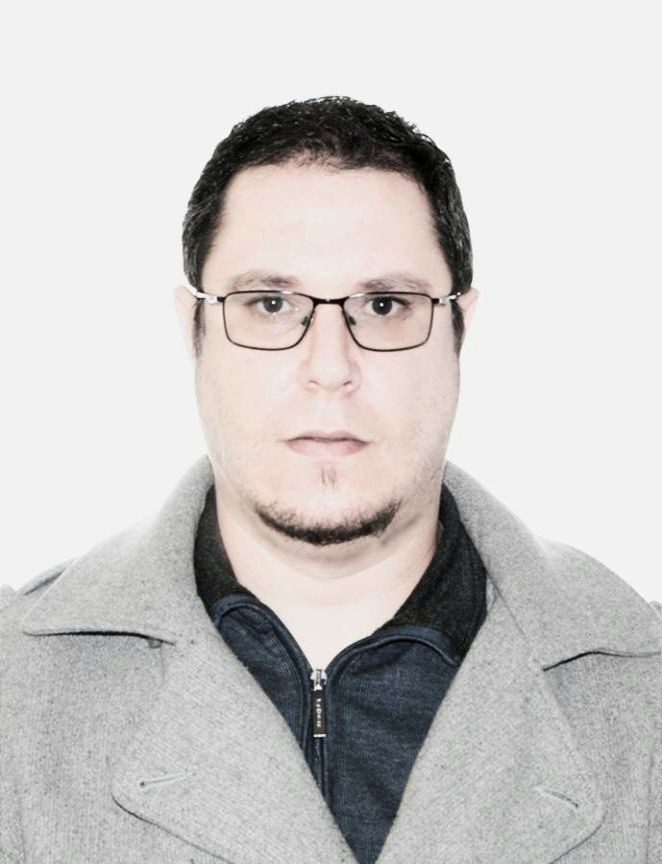}}] {Amine Boukhtouta} is an Experienced Researcher at Ericsson Security Research Group. He received the computer science engineering degree from USTHB University, Algiers, Algeria, in 2005 and the Master of Applied Science degree in information systems security degree and the Ph.D. degree in electrical and computer engineering from Concordia University, Montreal, Canada, in 2009 and 2016, respectively. He was part of Cyber-Forensics Training Alliance Canada, doing research on the generation of cyber-threat intelligence based on malware and network traces. He joined a Post-doctoral industrial program in 2016, where he worked on finding malicious indicators in evolving delivery network by applying big data analytics and machine learning. His current research interests include prevention, detection of cyber-threats by applying machine learning, and artificial intelligence. He published 5 journal papers and 11 conference papers in peer-reviewed venues. He was a recipient of OCTAS Prize in 2009 University Competition, the FQRNT Doctoral Scholarship in 2010–2011, the Best Paper Award, and the MITACS as well as PROMPT Postdoctoral Fellowships in 2016–2017. He is an expert in the Ericsson cybersecurity team.
\end{IEEEbiography}

\vskip -2pt plus -1fil

\begin{IEEEbiography}[{\includegraphics[width=1in,height=1.25in,clip,keepaspectratio]{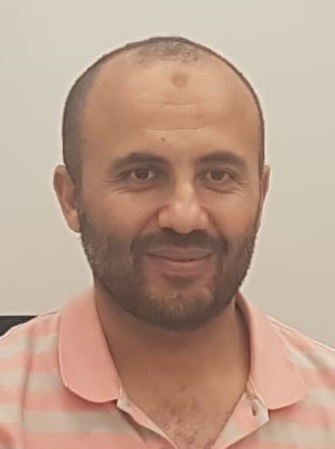}}] {Adel Larabi} is a Senior Solution Architect at Ericsson with over 25 years of leadership experience in designing innovative business solutions for Telco. Helping bridging academia research projects with commercial grade enterprise solutions. Core qualifications in CDN, Edge Computing, Big data, IMS, Media, and OSS with interest on AI applied to these domain. He is an expert in the Ericsson cybersecurity team.
\end{IEEEbiography}

\vskip -2pt plus -1fil

\begin{IEEEbiography}[{\includegraphics[width=1in,height=1.25in,clip,keepaspectratio]{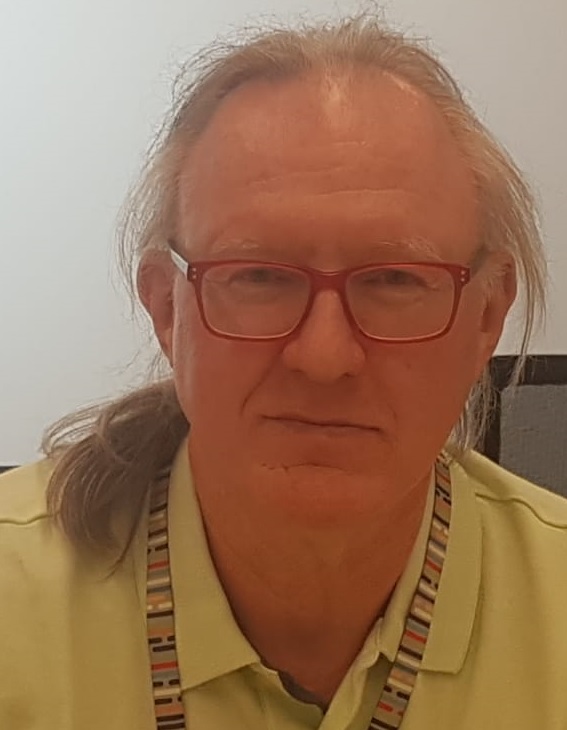}}] {Richard Brunner} joined Ericsson in 1988 and with broad career experience in System Management, Standardization and Strategic Product management. Richard possesses international management experience, with deep knowledge of the wireless telecommunication market. Richard is actively engaged in setting Ericsson’s research and partnership activities both internally and externally towards industry and universities. He is an expert in the Ericsson cybersecurity team.
\end{IEEEbiography}

\vskip -2pt plus -1fil

\begin{IEEEbiography}[{\includegraphics[width=1in,height=1.25in,clip,keepaspectratio]{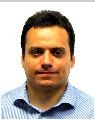}}] {Stere Preda} received his PhD in Computer Science from TELECOM Bretagne, France. Senior Researcher with expertise in cybersecurity at Ericsson, he is an active contributor to ETSI NFV security standardization.
\end{IEEEbiography}

\vskip -2pt plus -1fil

\begin{IEEEbiography}[{\includegraphics[width=1in,height=1.25in,clip,keepaspectratio]{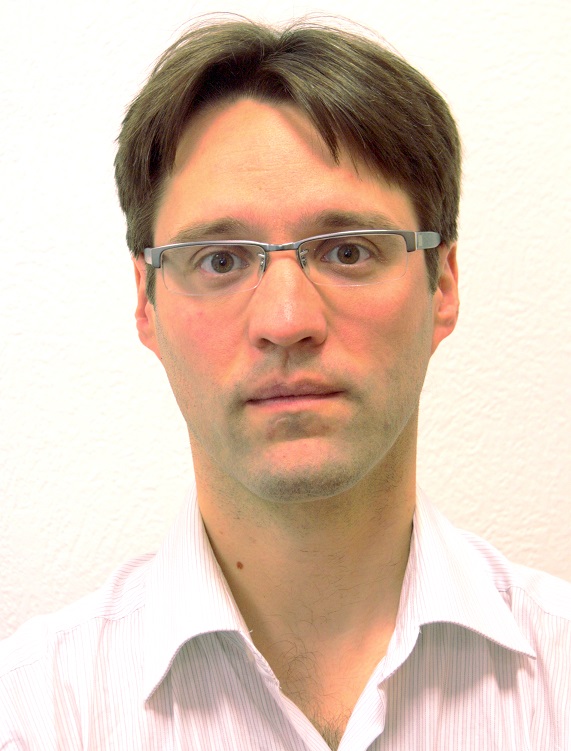}}] {Daniel Migault} is an expert in the Ericsson cybersecurity team and is actively involved in standardizing security protocols at the IETF.  
\end{IEEEbiography}

\end{document}